\pdfoutput=1 
\newif\ifarXiv
\arXivtrue 

\ifarXiv
\documentclass[aps,pre,reprint,superscriptaddress]{revtex4-2}

\else

\fi

\usepackage{graphicx}\graphicspath{{images/}} 
\usepackage{amsmath,amssymb,amsfonts,bm,siunitx,mathtools} 
\usepackage{hyperref} 
\usepackage{color}\definecolor{darkblue}{rgb}{0,0,0.5} 
\hypersetup{bookmarks=true, unicode=false, pdftoolbar=true, pdfmenubar=true,
pdffitwindow=false, pdfstartview={FitH}, pdfcreator={}, pdfproducer={},
pdfkeywords={} {} {}, pdfnewwindow=true, colorlinks=true, linkcolor=red,
citecolor=darkblue, filecolor=magenta, urlcolor=darkblue,}
\usepackage{cleveref} 
\usepackage{macros_quoting} 
\usepackage[mathlines]{lineno}
\usepackage{microtype}
\usepackage{placeins} 

\newcommand{\nn}{\nonumber} 
\newcommand{\E}{\mathop{{}\mathbb{E}}}

\newcommand{\ie}{{i.e.},~}
\newcommand{\eg}{{e.g.}\ }
\newcommand{\fig}[1]{Fig.~\ref{fig:#1}}

\newcommand{\secRef}[1]{Section~\ref{sec:#1}}

\newcommand{\eq}[1]{Equation~(\ref{eq:#1})}
\newcommand{\paperTitle}{A mathematical perspective on edge-centric brain functional connectivity}

\begin{document}

\ifarXiv

\title{\paperTitle}


\author{Leonardo Novelli}
\email[]{leonardo.novelli@monash.edu}
\affiliation{Turner Institute for Brain and Mental Health, School of Psychological Sciences and Monash Biomedical Imaging, Monash University, Australia}
\author{Adeel Razi}
\affiliation{Turner Institute for Brain and Mental Health, School of Psychological Sciences and Monash Biomedical Imaging, Monash University, Australia}
\affiliation{Wellcome Centre for Human Neuroimaging, University College London, United Kingdom}
\affiliation{CIFAR Azrieli Global Scholars Program, CIFAR, Toronto, Canada}


\date{\today}

\fi

\begin{abstract}
\textls[15]{Edge time series are increasingly used in brain functional imaging to study the node functional connectivity (nFC) dynamics at the finest temporal resolution while avoiding sliding windows.
Here, we lay the mathematical foundations for the edge-centric analysis of neuroimaging time series, explaining why a few high-amplitude cofluctuations drive the nFC across datasets.
Our exposition also constitutes a critique of the existing edge-centric studies, showing that their main findings can be derived from the nFC under a static null hypothesis that disregards temporal correlations.
Testing the analytic predictions on functional MRI data from the Human Connectome Project confirms that the nFC can explain most variation in the edge FC matrix, the edge communities, the large cofluctuations, and the corresponding spatial patterns.
We encourage the use of dynamic measures in future research, which exploit the temporal structure of the edge time series and cannot be replicated by static null models.}
\newpage
\end{abstract}

\ifarXiv
\maketitle
\fi


\section{Introduction}
\q{R1-introduction}{
Functional connectivity (FC) refers to patterns of statistical dependence in brain activity, such as the blood oxygen level-dependent (BOLD) signal measured via functional magnetic resonance imaging (fMRI).
Static FC is traditionally calculated over the course of an entire scan session and it is an established technique of modern neuroimaging~\citep{Craddock2013,Rogers2007}, with individual differences linked to brain disorders and cognitive states~\citep{Fornito2015,Cole2014}.
On the other hand, time-varying FC refers to time-resolved fluctuations in FC, typically estimated by fitting dynamic models~\citep{Park2018dynamic,Kashyap2019dynamic,Heitmann2018putting,Hansen2015} or by sliding windows~\citep{Sakoglu2010AMethod,Chang2010TimeFrequency}.
Differences in time-varying FC at rest are also associated with a wide range of cognitive and behavioural traits, as well as psychiatric and neurological conditions (see~\citep{Lurie2020} for a recent review of this rapidly growing field).

To avoid the temporal blurring caused by sliding windows~\citep{Sakoglu2010AMethod}, it is possible to analyse BOLD fluctuations at the resolution of single frames.
Coactivation patterns (CAPs)\citep{Liu2013} and point-process analyses~\citep{Tagliazucchi2012} are early examples of this approach.
Initially using a single seed region, both found that that static seed-based FC maps can be reliably approximated by averaging only a few high-amplitude frames.
Whole-brain extensions of both methods have soon followed~\citep{Liu2013Decomposition,Tagliazucchi2016}.
More recently, edge-centric approaches have generated excitement as they also go beyond single seeds and additionally decompose the entire FC matrix into its frame-wise contributions by omitting traditional time averaging over the length of the experiment~\citep{Esfahlani2020,Faskowitz2020}.
Such temporal unwrapping of the FC results in a large number of edge time series, each capturing the moment-to-moment cofluctuations of a pair of brain regions.
Going one step further, one can measure the similarity between every pair of edges to obtain a large matrix, referred to as edge FC (eFC), as opposed to the traditional node-centric FC (nFC).
The eFC is shown to be replicable, stable within individuals across multiple scanning sessions and reliable across datasets~\citep{Faskowitz2020}.
Furthermore, clustering the eFC yields overlapping brain communities that could better suit the study of aspects of cognition and behaviour that transcend traditional disjoint brain parcellations.
While CAPs focus on patterns of BOLD activity, edge-centric approaches focus on patterns of cofluctuations between all regional pairs.
However, the analogous finding is that the static FC can be faithfully approximated by averaging a few instantaneous connectivity patterns.
These are characterised by simultaneous large cofluctuations across all node pairs, measured as the root-sum-of-squares (RSS) over all the edge time series values corresponding to the same frame.
Such brief, intermittent, and high-amplitude cofluctuations drive the nFC and the network structure over these time points contributes disproportionately to the overall modularity of the functional brain network~\citep{Esfahlani2020}.
}

Although the edge-centric decomposition of nFC into its frame-wise constituents is mathematically exact, a comprehensive treatment of the statistical properties of the edge-centric measures is lacking and there is a consensus on the need for appropriate null models~\citep{Faskowitz2020,Betzel2021}.
A rigorous mathematical study is especially important since several widely-acknowledged publications have warned about the dangers of extracting structure from noise when studying static or time-varying FC, often using minimal null models to reproduce existing results~\citep{Liegeois2017,Lindquist2014,Laumann2017,Hindriks2016,Hlinka2015,Zalesky2012}.
The warnings concerning sampling variability are particularly relevant to edge-centric methods as they represent an extreme case of a single-frame sliding window approach.
High-amplitude cofluctuations can be observed in temporally-uncorrelated synthetic time series such that accounting only for static spatial correlations is sufficient to replicate key empirical findings~\citep[Fig. S4]{Esfahlani2020}.
This observation has been interpreted as further evidence that large cofluctuations are not fMRI artefacts.
However, it arguably raises an equally pressing conceptual concern: what information do current edge-centric measures provide beyond the nFC, if any?

Here, we tackle this question mathematically and present a theoretical explanation for the widespread occurrence of large cofluctuations across datasets and why a few large events drive nFC.
This explanation rests on fundamental properties of subexponential distributions~\citep{Foss2013}.
Further mathematical derivations clarify how the nFC eigenvalues shape the RSS distribution and how the leading nFC eigenvectors underpin the spatial correlation patterns expressed during high-amplitude events.
The influence of functional modules on the eigenvalue distribution could explain why these events disappear when the modular structure is disrupted, as recently reported in~\citep{Pope2021}.
Finally, we analytically show that the eFC matrix, the edge communities, the large cofluctuations, and the corresponding brain activity modes can all be predicted from the nFC without recourse to the edge-centric formulation.
Many of these derivations are based on the null hypothesis of i.i.d. Gaussian variables that only takes into account the observed (static) spatial correlations and ignores temporal features.
Under this assumption, and invoking results from random matrix theory~\citep{Wigner1967}, the edge time series variability is described by the sampling distribution of the nFC, known as the Wishart distribution~\citep{Wishart1928}.
Testing the analytic predictions using fMRI data from the Human Connectome Project (HCP)~\citep{HCP} shows that the null model is sufficient to replicate the vast majority of existing edge-centric features both qualitatively and quantitatively, as well as foundational properties of CAPs.

\section{Results}
\label{sec:results}
\begin{figure*}
    \ifarXiv\includegraphics[width=\textwidth]{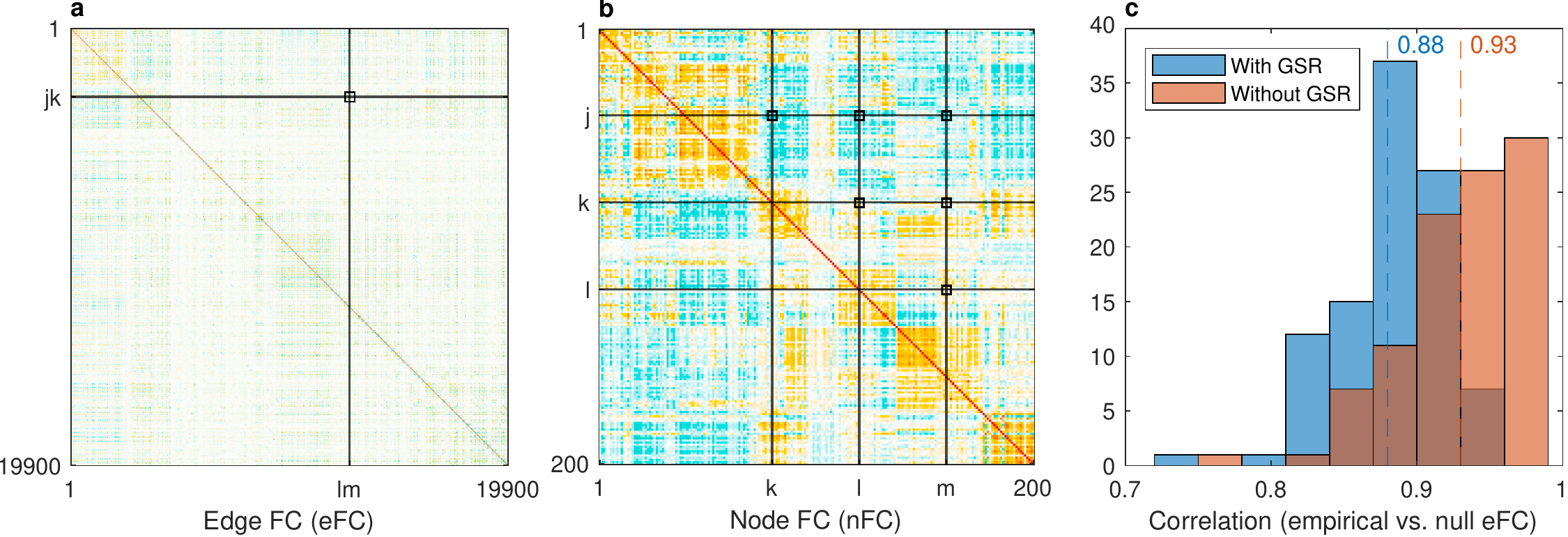}
    \else\centering\includegraphics[width=\textwidth]{eFC_from_nFC_GSR}
    \fi
    \caption{\label{fig:eFC_from_nFC_GSR}
        The edge functional connectivity (eFC) can be derived analytically from the node functional connectivity (nFC) under the static Gaussian null model.
        \textbf{a)} The eFC of the first subject from the Human Connectome Project (HCP) fMRI dataset is plotted as an example.
        This matrix scales with the fourth power of the parcellation size: in the current case of $200$ parcels, it has approximately $400$ million elements.
        The $(jk,lm)$ entry of the eFC matrix (black square at the intersection of the black lines) is computed as the inner product of two edge time series $jk$ and $lm$, normalised to the $[-1,1]$ interval.
        \textbf{b)} Under the static null model, the eFC of each subject can be derived analytically as a sum of pairwise products between the $(j,k,l,m)$ entries of the corresponding nFC matrix as per \eq{eFC_null_results}.
        Since the nFC is symmetric, only six entries are necessary to represent all pairwise combinations (marked by each of the black squares at the intersections of the black lines).
        \textbf{c)} Distribution of Pearson correlation coefficients between the empirical and null eFC computed over $100$ HCP subjects.
        Results are reported with and without global signal regression (GSR) and the mean correlation values are indicated by the dashed lines.
    }
\end{figure*}
We present six main results showing that the existing findings based on edge time series~\citep{Esfahlani2020,Faskowitz2020} can be derived from the static nFC under the null hypothesis of i.i.d. multivariate Gaussian variables that preserve the observed static spatial correlations but not the temporal ones.
The theoretical predictions were empirically tested using a HCP dataset comprising $100$ subjects, preprocessed with the current standard HCP pipeline, both with and without global signal regression (GSR).
The detailed derivations of the presented equations are in the Methods section, reserving the current section for a concise account of the key results only.

\subsection{The edge FC matrix can be derived analytically from the node FC}
The eFC was introduced in~\citep{Faskowitz2020} to quantify linear interactions between edges.
For each pair of brain regions, an edge time series is computed as the element-wise product of the two regional z-scored time series.
Thus, the values of each edge time series represent the instantaneous cofluctuation magnitudes between the corresponding pair of brain regions.
The eFC is the edge-by-edge matrix obtained by computing the inner products between all pairs of edge time series, normalised to the $[-1,1]$ interval.

Our first result is that the eFC can be analytically derived from the nFC under the static Gaussian null hypothesis (\fig{eFC_from_nFC_GSR}a,b).
As shown in \Crefrange{eq:ets_product}{eq:eFC_null} of the Methods section, the $(jk,lm)$ entry of the eFC matrix is obtained as a sum of pairwise products between the $(j,k,l,m)$ entries of the nFC matrix, divided by a normalisation factor:
\begin{align}\label{eq:eFC_null_results}
    \mathrm{eFC}_{jk,lm} =& \frac{r_{jk}r_{lm} + r_{jl}r_{km} + r_{jm}r_{kl}}{\sqrt{1 + 2 {r_{jk}}^2}\sqrt{1 + 2 {r_{lm}}^2}}.
\end{align}
Using the HCP dataset, the predicted eFC achieves an average Pearson correlation of $r=0.93$ with the empirical eFC ($r=0.88$ if GSR is applied).
The distributions across $100$ unrelated HCP subjects are shown in \fig{eFC_from_nFC_GSR}c.
This is a significant improvement on the linear regression approach adopted in~\citep{Faskowitz2020}, which achieved an average Pearson correlation of $r=0.72$ on pairs of edges not sharing any nodes, but performed poorly otherwise ($r=0.06$).
Moreover, by revealing the mathematical relationship between eFC and nFC, \eq{eFC_null_results} explains why the eFC is highly replicable, stable within individuals across multiple scan sessions and consistent across datasets---it will be as long as the nFC is.

\subsection{The edge communities can be predicted from the nFC}
\begin{figure*}
    \ifarXiv\includegraphics[width=0.7\textwidth]{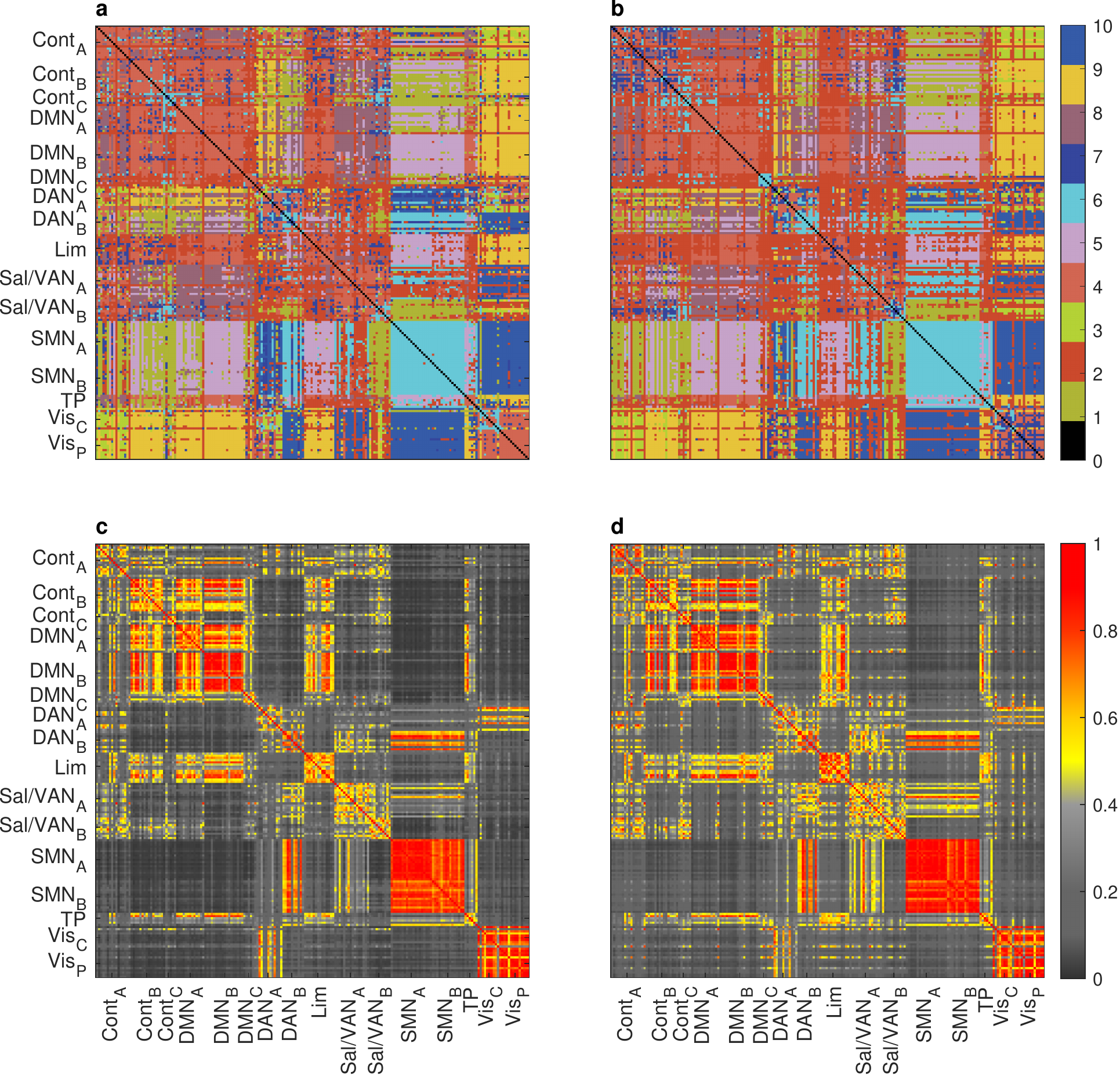}
    \else\centering\includegraphics[width=0.7\textwidth]{edge_communities_and_similarity_GSR}
    \fi
    \caption{\label{fig:edge_communities_and_similarity_GSR}
        The edge communities can be predicted from the nFC under the static Gaussian null model.
        \textbf{a)} Empirical edge communities obtained by k-means clustering of the eFC computed on the HCP dataset.
        The edge labels are reshaped into an $NxN$ matrix where each $(i,j)$ entry represents the community label of the edge linking $i$ to $j$.
        For each edge, the representative community assignment shown here is the statistical mode (\ie~the most common label) across $100$ HCP subjects.
        \textbf{b)} Predicted edge communities obtained by k-means clustering of the eFC predicted by the static null model.
        The agreement with the empirical labels in panel \textbf{a} is exact on $84\%$ of all $19900$ edges ($74\%$ if GSR is applied).
        \textbf{c)} Empirical edge cluster similarity on the HCP dataset.
        The similarity of edge communities involving nodes $i$ and $j$ is computed as the fraction of matching elements between the corresponding two rows of the edge community matrix in panel \textbf{a}.
        \textbf{d)} Predicted edge cluster similarity based on the static null model.
        The Pearson correlation with the empirical edge cluster similarity in panel \textbf{c} is $r=0.96$ ($r=0.95$ if GSR is applied).
    }
\end{figure*}
Clustering the eFC via the k-means algorithm was used to identify communities of co-fluctuating edges.
These were then mapped back to individual nodes to obtain overlapping regional communities, offering a new way to study aspects of cognition and behaviour that transcend traditional disjoint brain parcellations~\citep{Faskowitz2020}.
Given the high similarity between the empirical and null eFC demonstrated in the previous section, it would be reasonable to expect that the edge communities be well recovered by the null model by applying the same community detection algorithm to the predicted eFC.
Indeed, the agreement between the empirical and predicted community assignments is exact on $84\%$ of the $19900$ edges, and $74\%$ if GSR is applied (\fig{edge_communities_and_similarity_GSR}a,b).
At the node level, the similarity is even stronger.
In~\citep{Faskowitz2020}, the similarity between two nodes, referred to as ``edge cluster similarity'', is measured as the fraction of all edge pairs starting from those two nodes (and reaching the same target) that are clustered together.
Using the same procedure, the predicted edge cluster similarity achieves a Pearson correlation coefficient of $r=0.96$ with the empirical one, and $r=0.95$ if GSR is applied (\fig{edge_communities_and_similarity_GSR}c,d).
Note that the rows and columns of the matrices in \fig{edge_communities_and_similarity_GSR} have been rearranged to match the ordering of the $16$ networks used in~\citep[Fig.6]{Faskowitz2020} to facilitate a visual comparison.
Given that the analysis is performed on the same HCP dataset, the small differences in the empirical results are most likely due to the different preprocessing pipelines (here we use the preprocessed data made available by the HCP; see the Methods section for details).
This adds to the evidence that edge-centric measures are not strongly dependent on the preprocessing approach, and similarly speaks to the robustness of the null model predictions.

So far, the results in this subsection have been based on simulations.
Let us now see how a mathematical derivation can provide further insight into these edge and node similarities.
First, recall that the edge communities are obtained by clustering the eFC (\eg~via k-means).
The main obstacle to a full analytic approach is that the outcome of stochastic clustering algorithms cannot be entirely predicted from their input; however, since they are usually based on a distance metric, it is reasonable to expect that the smaller the distance between two rows of the eFC, the higher the probability that the corresponding edges would be clustered together.
Accordingly, we define the distance between two edges $(jk)$ and $(j'k')$ as the $\ell^1$ norm of the difference between the corresponding rows of the eFC.
As shown in \eq{dist_edges} of the Methods section, this edge distance simplifies to
\begin{align} \label{eq:dist_edges:results}
    d_{jk,j'k'} =\sum_{l,m=1}^N &3|\mathbf{z}_j (\mathbf{z}^\intercal_l \mathbf{z}_m) \mathbf{z}^\intercal_k - \mathbf{z}_{j'} (\mathbf{z}^\intercal_l \mathbf{z}_m) \mathbf{z}^\intercal_{k'}|\ ,
\end{align}
where, for a brain region $i$, the row vector $\mathbf{z}_i$ is its z-scored BOLD signal and $\mathbf{z}^\intercal_i$ denotes its transpose.
What does this imply for the node communities?
The similarity between two nodes $i$ and $j$ was measured in \citep{Faskowitz2020} as the fraction of all edge pairs starting from $i$ and $j$ (and reaching the same target) that are assigned to the same cluster.
Here, the analogous analytic step is to compute the distance between nodes $i$ and $j$ as the sum of the distances between the edges starting from them.
Crucially, the resulting node distance can be expressed in terms of BOLD signal correlations (see \eq{dist_nodes} in the Methods section): 
\begin{align}
    d_{i,j} \leq c \left(1-r_{ij}\right)^\frac{1}{2},
\end{align}
where $c$ denotes a constant term, independent of $i$ and $j$.
This implies that the edge-cluster similarity~\citep{Faskowitz2020} between nodes $i$ and $j$ can be predicted from the corresponding $r_{ij}$ entry of the nFC, avoiding the memory-intensive computation of the eFC and computationally-intensive clustering algorithms (the space complexity of the eFC is $\mathcal{O}(N^4)$, \ie~it scales with the fourth power of the number of regions and requires over a terabyte of memory for fine brain parcellations---for each subject).
Using the HCP dataset to test this prediction shows that the nFC alone achieves an average Pearson correlation of $r=0.76$ with the empirical edge cluster similarity matrix shown in \fig{edge_communities_and_similarity_GSR}c.
\q{R2-finelytunedcofluctuations1}{Once again, the static, node-centric, second-order features of the BOLD signal are sufficient to replicate key findings that appear at first to rely on the specific temporal sequence of BOLD cofluctuations at the single-frame resolution.}

\subsection{The null model reproduces the high similarity of the top RSS frames to the nFC}
\begin{figure*}
    \ifarXiv\includegraphics[width=0.95\textwidth]{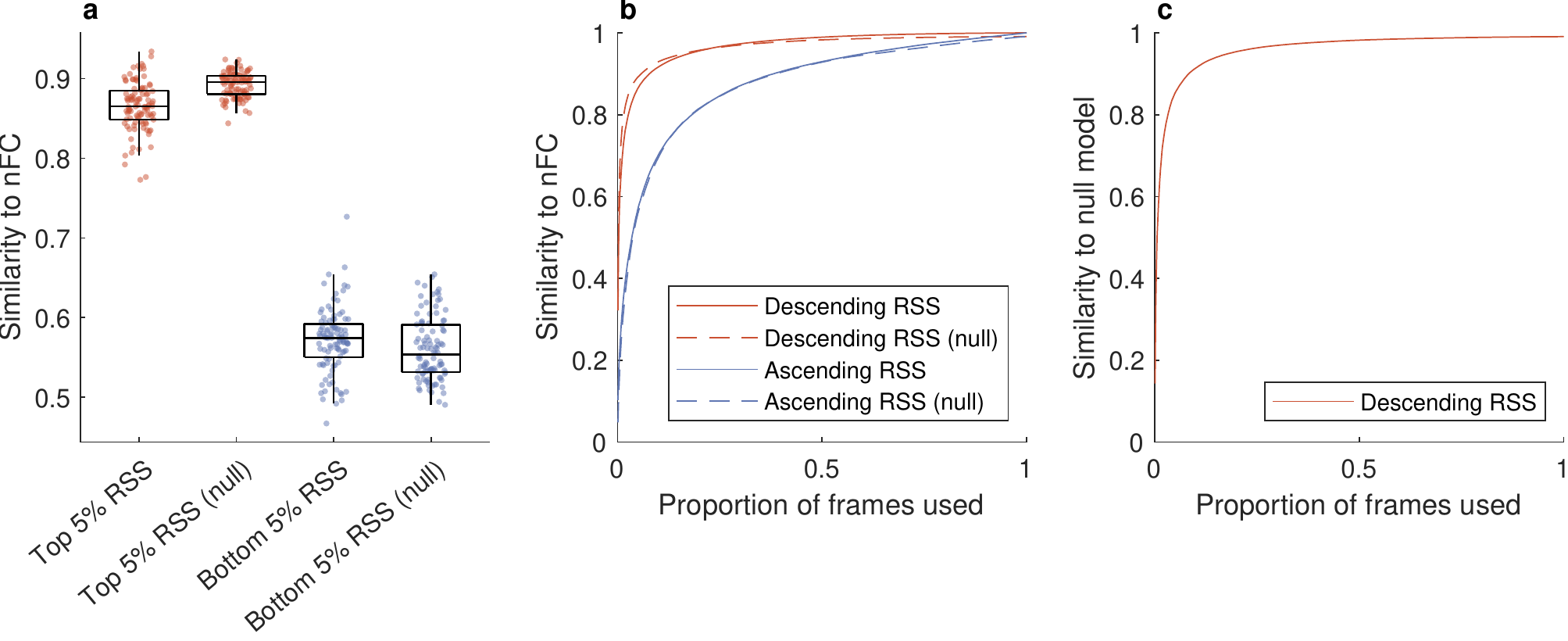}
    \else\centering\includegraphics[width=0.95\textwidth]{nFC_similarity_GSR}
    \fi
    \caption{\label{fig:nFC_similarity_GSR}
        The static Gaussian null model can reproduce the strong correlation between high-amplitude cofluctuation patterns and the nFC.
        \textbf{a)} Only a small fraction of frames exhibiting the largest cofluctuations root-sum-of-squares (RSS) are required to explain most of the nFC variance.
        The similarity is computed as the Pearson correlation coefficient between the nFC and the average FC estimated from the top and bottom $5$\% of the total frames.
        Each point corresponds to one of $100$ unrelated subjects from the Human Connectome Project (HCP) dataset, with boxes indicating the quartiles and whiskers length specified as $1.5$ times the interquartile range.
        \textbf{b)} The same results hold more generally when the frames are ordered according to the corresponding RSS amplitude, either in descending or ascending order.
        Here, the curves represent the average similarity over $100$ subjects.
        \textbf{c)} The findings do not depend on the timing of the high-amplitude RSS events: the top frames generated by the null model exhibit high similarity to the top frames of the real HCP data, which occur at different times.
    }
\end{figure*}

\q{R1-EuclideanNormRSS1}{
Let us now consider the root-sum-of-squares (RSS) of the edge time series introduced in~\citep{Esfahlani2020}.
The RSS is a univariate time series defined as the Euclidean norm of the edge time series vector at each frame.
In other words, the RSS peaks when all the cofluctuations (\ie~the edge time series) are simultaneously high in absolute value, either positive or negative.
}
The key finding in~\citep{Esfahlani2020} is that only a small fraction of frames exhibiting large RSS values are required to explain a significant fraction of variance in the nFC, as well as the network’s modular structure.
Both results are perfectly reproduced by the static null model, as shown in \fig{nFC_similarity_GSR}a,b and \fig{modularity_GSR}.
What is particularly remarkable is that the timing of the high-amplitude RSS events produced by the null model are arbitrary, and yet a small fraction of frames corresponding to these large cofluctuations is still sufficient to explain the observed nFC.
Furthermore, \fig{nFC_similarity_GSR}c shows that the null model frames with the largest RSS also exhibit high similarity to the empirical frames with the largest RSS from the HCP dataset---occurring at entirely different times.
Note that, unlike temporal dependencies, spatial correlations are necessary to reproduce the results: if the null model is chosen to be both temporally and spatially uncorrelated, high-RSS frames are no more similar to the empirical nFC than low-RSS frames (see \fig{nFC_similarity_white_noise_GSR}).
Another interesting note is that high-RSS frames exhibit the strongest average correlation with all other BOLD frames and this average similarity decreases with the RSS magnitude, both in empirical and null cases (see \fig{similarity_to_all_frames_vs_descending_RSS_subj5}).

A theoretical explanation for these findings will be provided in \secRef{spatial_mode} and supported by detailed derivations in the Methods section.
Interestingly, most of these points were also reported in \citep[Fig. S4]{Esfahlani2020}, where they were taken as evidence that large RSS events are not fMRI artefacts.
While settling that methodological issue, these observations raise a conceptual concern: if matching the timing of the RSS events is not essential and the results can be replicated by the static null model, does the edge-centric approach provide any information about the time-varying connectivity that cannot be explained by the static nFC?
We will address this question in the next section, by examining the statistical properties of the RSS.

\subsection{The RSS distribution is determined by the nFC eigenvalues}
\begin{figure*}
    \ifarXiv
    \includegraphics[width=0.76\textwidth]{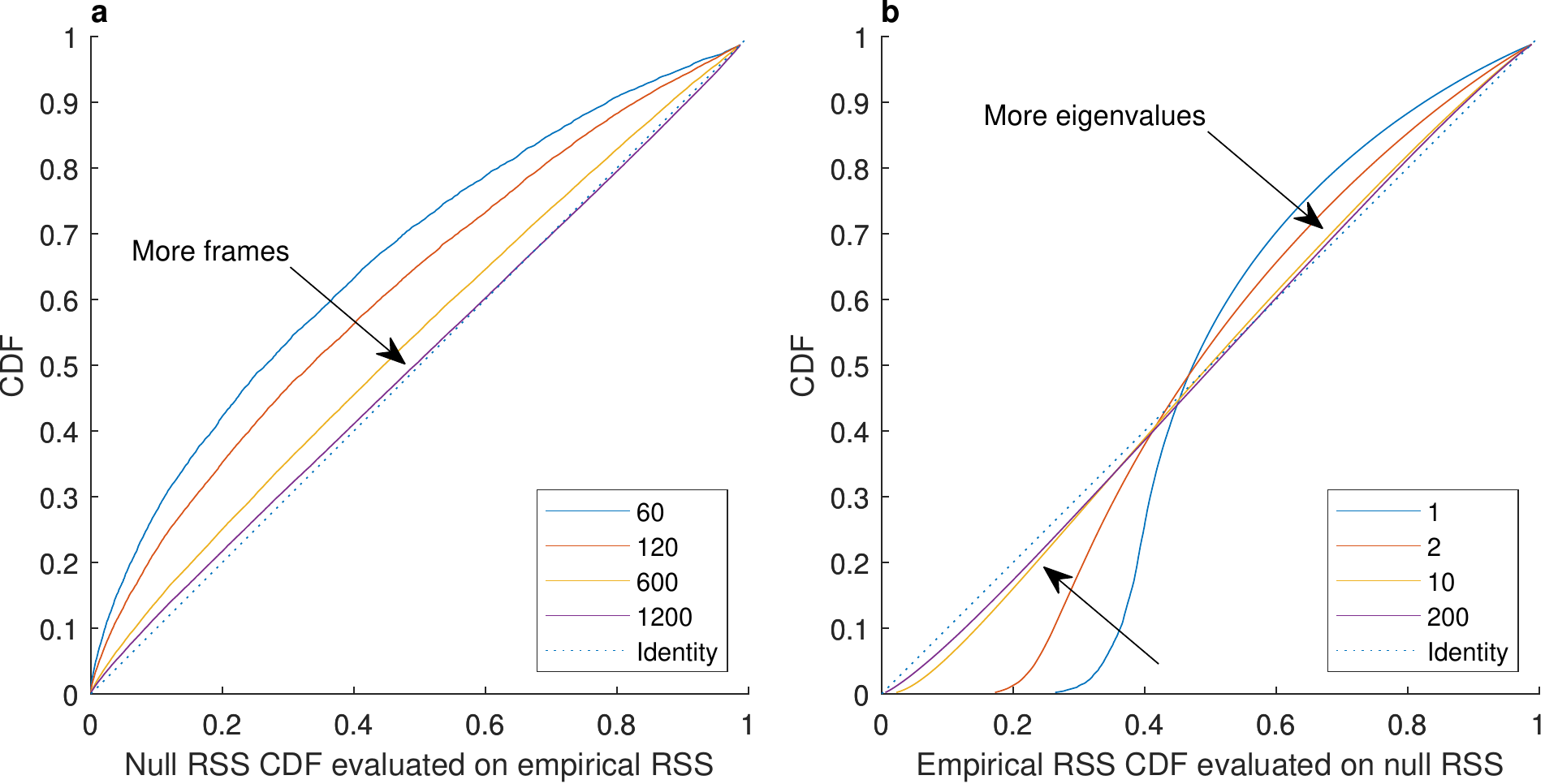}
    \else\centering
    \includegraphics[width=0.8\textwidth]{CDFCDF_plots_GSR}
    \fi
    \caption{\label{fig:CDFCDF_plots}
        Properties and convergence of the RSS distribution derived analytically under the static null hypothesis.
        \textbf{a)} The empirical RSS distribution increasingly approximates the null RSS distribution as more realisations (frames) are observed.
        On the horizontal axis, the cumulative distribution function (CDF) of the null RSS is evaluated for each empirical RSS observation; the CDF of the resulting distribution is plotted on the vertical axis.
        Identical distributions would produce points on the diagonal (dotted line).
        \textbf{b)} The nFC eigenvalues shape the null RSS distribution.
        The null RSS distribution increasingly approximates the empirical RSS distribution as more eigenvalues are considered, with the largest eigenvalues explaining most of the variance.
        On the horizontal axis, the cumulative distribution function (CDF) of the empirical RSS is evaluated for each null RSS observation; the CDF of the resulting distribution is then plotted on the vertical axis.
        Identical distributions would produce points on the diagonal (dotted line).
    }
\end{figure*}
Having established that the large RSS events are not an exclusive feature of neural signals, let us investigate how their ubiquitous appearance can be analytically explained and why the corresponding frames account for the largest fraction of variance in the nFC.
As a first step, the RSS can be computed as the squared Euclidean norm of the z-scored BOLD signal $\mathbf{z}$ (full derivations provided in the Methods section, see \Crefrange{eq:RSS_as_Euclidean_approx}{eq:RSS_as_Euclidean_exact}):
\begin{equation}
    \mathrm{RSS}_\mathrm{all}(t) ={\lVert \mathbf{z}(t)\rVert}^2.
\end{equation}
The intuition behind this equivalence is that summing over the pairwise products of the elements of a vector is the same operation that is performed when squaring a polynomial; in this case, the vector is the squared BOLD signal, its pairwise products are the squared edge time series, and the polynomial is the squared Euclidean norm.
\q{R1-EuclideanNormRSS2}{The key message is that, although it was introduced in~\citep{Esfahlani2020} as the Euclidean norm of the edge time series, the RSS is mathematically equivalent to the squared Euclidean norm of the BOLD signal, \ie~a measure of the fluctuation of the BOLD signal amplitude over time.}
We can then proceed without resorting to the edge time series, which is not only convenient in practice but also shifts the conceptual focus back to the BOLD time series---which are more readily interpretable.
For a large family of common (sub-Gaussian) distributions, the squared Euclidean norm of a random variable (RV) is heavy-tailed (more specifically, it is sub-exponential~\citep{Wainwright2019}).
The RSS being a squared norm, large cofluctuations are then to be expected due to its heavy-tailed distribution (see \eq{sub_gaussian}), offering an explanation for the large RSS peaks observed in the BOLD time series.

In the specific case of Gaussian variables (\ie~the null hypothesis), the RSS can be expressed as a sum of $N$ independent Gamma($k=1/2,\theta=\sqrt{2}\lambda_i$) variables, each related to an eigenvalue of the nFC matrix ($\lambda_i,i=1\ldots,N$).
This is summarised by the moment-generating function in \eq{RSS_mgf}.
The largest eigenvalues capture the distribution tail and including smaller eigenvalues provides an increasingly complete characterisation of the empirical RSS distribution (\fig{CDFCDF_plots}b).
This distribution can be used for testing the statistical significance of the empirical RSS observed in the HCP dataset against the static null hypothesis of spatially correlated noise.
\fig{CDFCDF_plots}a illustrates the convergence of the empirical RSS distribution to the null distribution as more and more time frames are observed (\ie~over longer fMRI sessions). 
When all the $1200$ time frames available in the HCP data are utilised, the null hypothesis cannot be rejected for $58$\% of the participants at a $5\%$ significance level and on $90$\% of the participants after Bonferroni correction for multiple comparisons (p-values given by the two-sided Kolmogorov-Smirnov test).

\subsection{nFC eigenvectors underpin spatial patterns of high BOLD activity}
\label{sec:spatial_mode}
\begin{figure*}
    \ifarXiv
    \includegraphics[width=\textwidth]{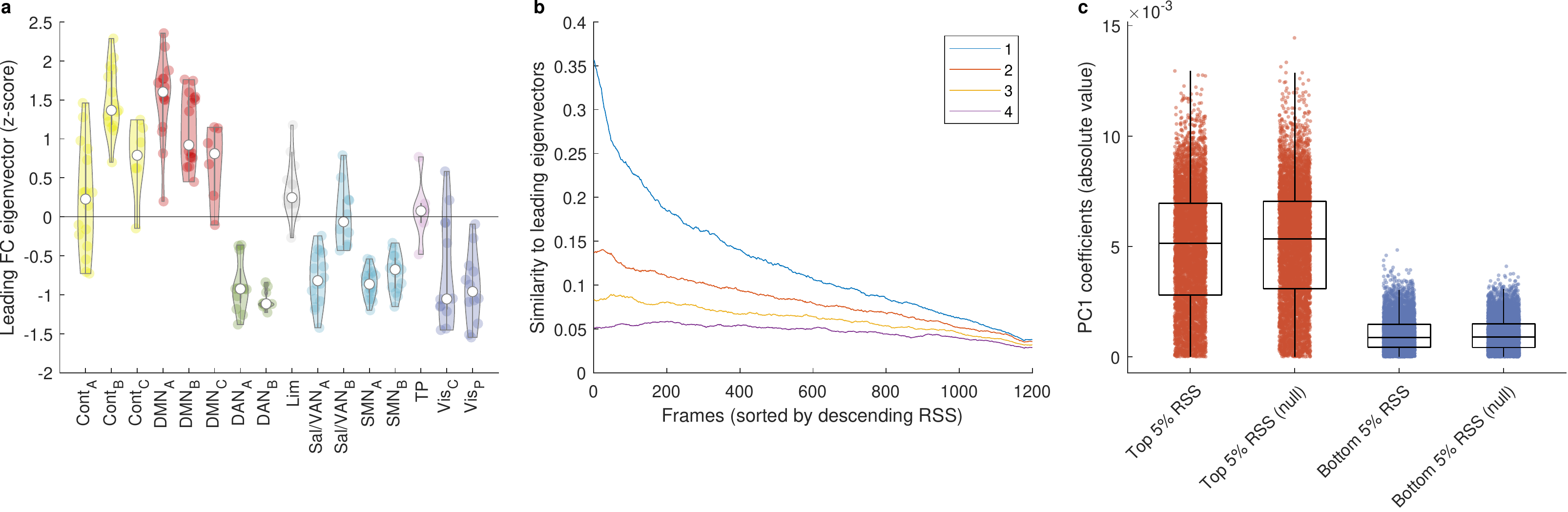}
    \else\centering
    \includegraphics[width=\textwidth]{PC_plots_GSR}
    \fi
    \caption{\label{fig:PC_plots_GSR}
        The static Gaussian null model can explain the spatial patterns of BOLD activity and the cofluctuation patterns that characterise high-amplitude frames.
        \textbf{a)} The spatial mode underpinning high-amplitude cofluctuations is captured by the leading eigenvector of the nFC matrix.
        The $200$ brain regions were partitioned into the same $16$ networks used in~\citep[Fig.2E]{Esfahlani2020} to facilitate a visual comparison.
        In addition to a scatter plot of the data, the violin plots show the probability densities and the box plots indicate the quartiles, with the maximum whisker length specified as $1.5$ times the interquartile range.
        The sample sizes of the box plots are the sizes of the $16$ networks, that is, $n=[16,15,6,14,17,6,12,10,14,16,10,19,15,6,12,12]$.
        \textbf{b)} Frames corresponding to large RSS values exhibit high similarity to the leading nFC eigenvectors.
        The similarity is measured as the Pearson correlation between the instantaneous FC estimate from a single frame and each of the estimates from the four leading nFC eigenvectors.
        \textbf{c)} Higher principal component (PC1) coefficients are associated with large RSS events.
        In addition to a scatter plot of the data, the box plots indicate the quartiles and the whisker length is specified as $1.5$ times the interquartile range (sample size $n=6000$).
        Compare this figure with the results published in~\citep[Fig.2C]{Esfahlani2020}.
    }
\end{figure*}

High-amplitude BOLD cofluctuations were observed to be underpinned by a particular spatial mode of brain activity in which default mode and control networks are anticorrelated with sensorimotor and attentional systems~\citep{Esfahlani2020}.
This particular spatial mode was defined as the first principal component of the BOLD activity and, as such, it can be obtained as the largest eigenvector of the static nFC matrix by mathematical equivalence and without recourse to null models (\fig{PC_plots_GSR}a).

The only question left to answer is whether the RSS is predicted to peak when the BOLD activity aligns with the largest eigenvector.
This can be proven to be true even beyond the static null hypothesis (see \Crefrange{eq:BOLD_aligned_with_PC1}{eq:ets_aligned_with_PC1} in the Methods section).
An intuitive understanding can be gained once the RSS is seen as the fluctuation of the BOLD signal amplitude over time: high-amplitude frames have a larger variance, which is captured by a larger coefficient of the first principal component since the latter is the vector that aligns with the direction of maximum variance (\fig{PC_plots_GSR}c).
We can then refine our theoretical understanding of the RSS peaks: not only do they occur when the Euclidean norm of the BOLD signal vector is large but, most likely, when it is well aligned with the leading eigenvector of the nFC.
However, the alignment with the leading eigenvector need not be perfect: in general, large RSS values can be expected whenever the BOLD vector is a mixture of the top eigenvectors (\fig{PC_plots_GSR}b and \fig{eigenvectors_vs_RSS_GSR_moreframes}).
Additional principal components are included as more large-RSS frames are averaged, suggesting why the top $5\%$ frames alone are sufficient for an almost perfect reconstruction of the nFC~(\fig{nFC_similarity_GSR}).   
We have thus explained why the nFC estimates corresponding to frames with the largest RSS exhibit the highest similarity with the nFC.
Moreover, since the nFC features multiple communities, high-RSS frames naturally reflect this property by exhibiting high modularity, as well as higher values than low-RSS frames, which are less similar to the nFC (see \fig{modularity_GSR}).

One could speculate that the cofluctuation patterns corresponding to large nFC eigenvectors would be closely related to empirical cofluctuation patterns obtained by clustering high-RSS frames as in~\citep{Betzel2021}.
For example, the leading nFC eigenvector has the highest similarity to all BOLD frames and the corresponding cofluctuation pattern would resemble the most frequently occurring cluster.

\subsection{A null model for binary edge time series}
It was recently observed that thresholding the edge time series retains most of the information about the nFC since averaging the binarised edge time series over time still yields a very accurate approximation of the nFC both at the voxel~\citep{Tagliazucchi2016} and parcel level~\citep{Sporns2021}.
As a premise, the latter finding is empirically replicated here using the HCP dataset, with a resulting Pearson correlation $r=0.98$ between the average binarised edge time series and the nFC (average correlation over $100$ unrelated subjects).
How to explain this almost perfect correlation?
Furthermore, it was noted that the binary edge time series are highly constrained by the nFC~\citep{Sporns2021}.
What is the nature of such constraints?
Both questions can be answered mathematically under the static Gaussian null hypothesis.
Consider two brain regions or parcels $j$ and $k$ with z-scored BOLD activity $Z_j(t)$ and $Z_k(t)$ and their corresponding edge time series $C_{jk}(t)=Z_j(t)Z_k(t)$.
The thresholded edge time series $\overline{C_{jk}(t)}$ is equal to $1$ when the cofluctuations between $j$ and $k$ are positive, \ie~when $Z_j(t)$ and $Z_k(t)$ have the same sign.
Under the null hypothesis, $Z_j(t)$ and $Z_k(t)$ are normal Gaussian RVs with correlation coefficient denoted as $r_{jk}$.
If the two parcels are uncorrelated ($r_{jk}=0$), their sign is positive or negative with equal probability and can be described mathematically as a Bernoulli($1/2$) RV, which is a formal way to model a fair coin draw.
The corresponding binary edge time series $\overline{C_{jk}(t)}$ can also be modelled as a fair coin draw since the two signs are expected to agree half of the times.
The situation clearly gets more complicated in the case of $200$ correlated parcels as in the HCP dataset considered here.
Intuitively, we would expect $\overline{C_{jk}(t)}=1$ if the two parcels were perfectly correlated and $\overline{C_{jk}(t)}=0$ if they were perfectly anticorrelated, \ie~in consistent disagreement.
This intuition can be formalised mathematically and extended to all intermediate cases as shown in \eq{bernoulli_p_methods} of the Methods section.
A trigonometric argument proves that $\overline{C_{jk}(t)}$ can be modelled as a biased coin, \ie~a Bernoulli($p_{jk}$) RV with success probability
\begin{align}\label{eq:bernoulli_prob_results}
    p_{jk}=\frac{1}{2}+\frac{\arcsin(r_{jk})}{\pi}.
\end{align}
This result reveals the exact analytic relationship between the binary edge time series and the nFC, answering the second question.
However, it is straightforward to show that \eq{bernoulli_prob_results} also predicts the approximated nFC obtained by averaging the binary edge time series.
This follows from the fact that the time average of ergodic processes (including the null model) converges to their expectation, and that the expectation of a Bernoulli RV is equal to its success probability, $p_{jk}$.
Testing this analytic prediction on the HCP dataset confirms its accuracy (see \fig{bernoulli_prediction_GSR}b).
Finally, we can explain the strong correlation between the original and approximated nFC by noting that $p_{jk}$ in \eq{bernoulli_prob_results} is very close to $r_{jk}$ for small values of the correlation (see \fig{bernoulli_prediction_GSR}a), which are the most frequent ones.


\subsection{Relationship with coactivation patterns (CAPs)}
\begin{figure*}
    \ifarXiv\includegraphics[width=0.7\textwidth]{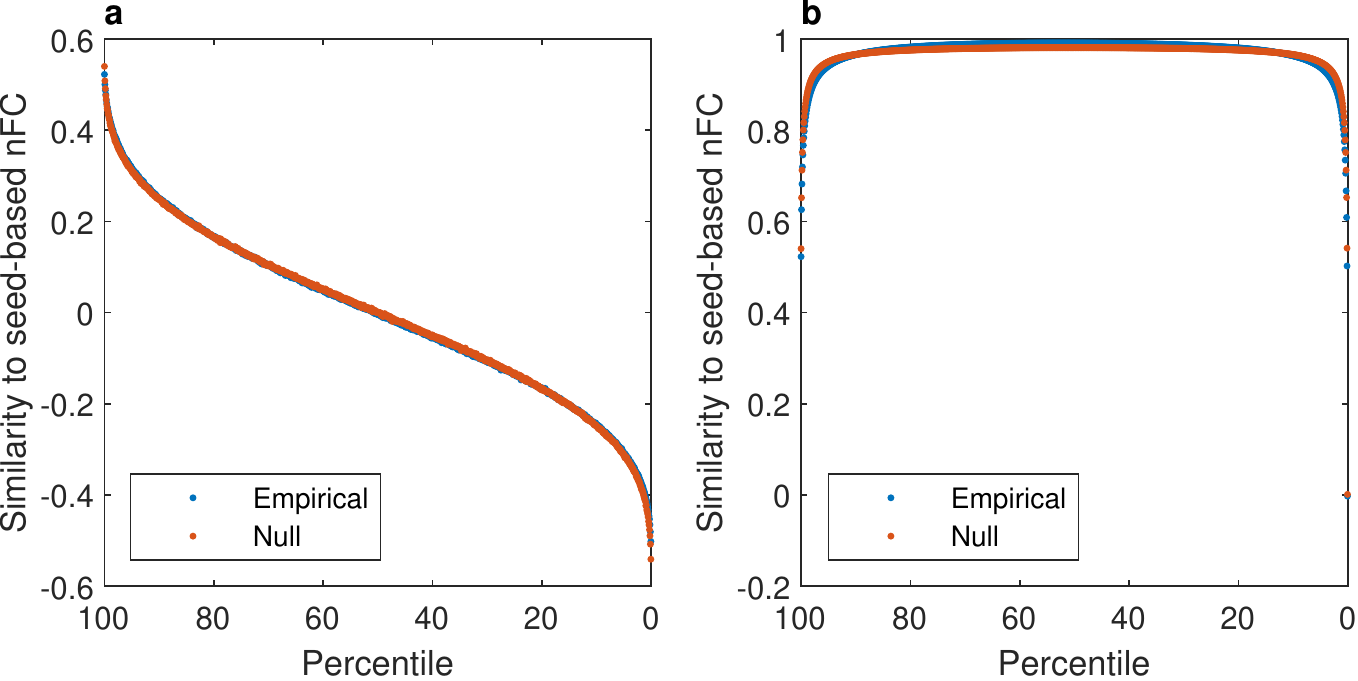}
    \else\centering\includegraphics[width=0.7\textwidth]{CAPs_similarity_GSR}
    \fi
    \caption{\label{fig:CAPs_similarity_GSR}
        The static Gaussian null model can explain the spatial similarity between high-amplitude frames based on a seed node and the corresponding correlation map---a core feature of coactivation patterns (CAPs).
        \textbf{a)} For each seed node (i.e. parcel), the frames are sorted in descending order based on the BOLD activity of the seed (horizontal axis).
        The Pearson correlation is computed between each frame and the seed-based correlation map (i.e.,~the column of the nFC matrix corresponding to the seed).
        The curves represent the average similarity over all $200$ seed nodes and over $100$ HCP subjects.
        As predicted analytically, the similarity is proportional to the BOLD activity and it decreases as lower-amplitude frames are considered (lower percentiles).
        The null model is in excellent agreement with the empirical results.
        \textbf{b)} Same as in panel \textbf{a}, but the BOLD activity of all the frames above a given threshold (percentile) are averaged before computing the correlation with the seed-based FC map.
        Only a small fraction of high-amplitude frames is required to explain most of the nFC variance, reproducing a well-known result in the CAPs literature~\citep{Liu2013}.
    }
\end{figure*}
CAPs~\citep{Liu2013,Liu2018} preceded edge-centric approaches in the study of spatial patterns of BOLD activity at the single-frame resolution.
These patterns differ from established resting-state functional networks and it was speculated that they may originate from a neuronal avalanching phenomenon.
Both CAPs and edge-centric studies report a strong similarity between high-amplitude frames and the nFC, and both employ k-means to assign these frames to different clusters~\citep{Liu2013,Betzel2021}.
However, there are important differences between the two methods.
CAPs are patterns of node coactivations (\ie~they are defined in node space), while the frame-wise FC patterns studied in edge-centric analyses are defined in edge space.
Moreover, CAPs are seed-based, \ie~frames to be clustered are selected based on high activity levels of a single node rather than simultaneous high activity across all nodes.
For example, choosing the posterior cingulate cortex (PCC) as the seed, \citet{Liu2013} observed a strong correlation between the BOLD activity at times where PCC is highly active and the corresponding seed-based FC map.

The static Gaussian null model employed in this study can predict this strong correlation, which is the fundamental conceptual and practical property guiding the selection of frames that are subsequently clustered into CAPs.
As shown in \eq{CAPS_expected_Z} of the Methods section, frames selected based on the high activity of a seed node are expected to exhibit BOLD activity patterns that reflect the corresponding column of the nFC.
Furthermore, the correlation is directly proportional to the activity level and thus peaks at frames with the highest activity of the seed.
This relationship is illustrated in (\fig{CAPs_similarity_GSR}a), where frames are sorted in descending order based on the seed node activity.
It is important to note that parcels are used instead of voxels, in alignment with the rest of the simulations herein; importantly, the mathematical results hold true in both cases.
What happens when more frames are averaged?
The answer is shown in \fig{CAPs_similarity_GSR}b and follows intuitively from \fig{CAPs_similarity_GSR}a.
Let us again consider the frames sorted in descending order; as the activity threshold is lowered (\ie~set to a lower percentile), more frames are averaged.
The first frames are best aligned with the seed FC vector (\ie~the nFC column corresponding to the seed node) and their average quickly converges to the same direction.
The alignment (or similarity) reaches a plateau as the middle frames are added to the average since they have low or zero correlation with the seed FC vector.
Finally, as the last and most negatively-correlated frames are added to the average, the similarity drops sharply.
This explains and replicates the first findings of the seminal CAPs paper by \citet{Liu2013}.

To summarise, analytic predictions in the case of CAPs differ from the edge-centric case in that the high-amplitude frames are best explained by individual nFC columns rather than its eigenvectors.
However, there is an intuitive link between the two: if we select frames where a specific node is highly active (CAPs approach), the BOLD signal will most likely resemble the corresponding nFC column; if we select frames where all nodes are highly active (edge-centric approach), the BOLD signal will most likely resemble all the nFC columns, \ie~it will align with the leading nFC eigenvector.

\section{Discussion}
\q{R2-finelytunedcofluctuations2}{
We have presented mathematical proofs and numerical analyses of real HCP data supporting our claim that the static nFC is sufficient to replicate the main resting-state edge-centric findings in~\citep{Esfahlani2020} and~\citep{Faskowitz2020} both qualitatively and quantitatively, without relying on the edge time series nor any temporal correlations.
}
Specifically, the eFC, the edge communities, the edge time series norm (RSS) distribution, and the spatial BOLD patterns underpinning large cofluctuations can all be predicted from the nFC under the null hypothesis of i.i.d. multivariate Gaussian variables.
The inability to reject the null hypothesis on most of the HCP $100$ unrelated subjects does not support the conclusion that these edge-centric metrics provide additional information beyond the nFC.
These results are not an attempt to disprove the existence of finely timed neural events---they just warn that the evidence provided by fMRI data may not be sufficient to reject simpler explanations, based on the established nFC literature, for the edge-centric features studied in~\citep{Esfahlani2020} and~\citep{Faskowitz2020}.
In fact, previous influential studies have raised similar warnings in the context of sliding-window approaches to time-varying FC~\citep{Liegeois2017,Laumann2017,Hindriks2016}.

However, it would be premature to conclude that the nascent edge-centric approach has no merit, and we acknowledge the fast progress in its development and applications at the time of writing~\citep{Betzel2021,Sporns2021,Pope2021}.
Particularly interesting is the influence of structural modules on the edge cofluctuations~\cite{Pope2021}, which we briefly address in the Methods section.
The size of the functional modules shapes the spectrum of the nFC: larger functional modules allow for larger eigenvalues, which underpin the high-amplitude cofluctuations.
This offers a mathematical insight into the relationship between modular structure and large cofluctuations, and why the latter disappear if the modular structure is disrupted.
While fully addressing the latest (and increasingly large number of) preprints is beyond the scope of this work, it is possible that model-based approaches will reveal the role of edge-centric properties in bridging brain structure and function.
Indeed, temporally-unfolded (or point-wise) dependence measures have been instrumental in studying the structure-function relationship in canonical complex systems~\citep{Lizier2008,Lizier2013}; seeing the edge time series as point-wise mutual information under the Gaussian assumption could create new links to the existing literature.

It would also be unreasonable to assume that the null hypothesis of i.i.d. variables is a good description of the BOLD signal, which is slowly-varying and highly autocorrelated.
Thus, the fact that such null model is able to replicate the edge-centric features in~\citep{Esfahlani2020} and~\citep{Faskowitz2020} could be an indication that the temporal structure of the edge time series has not been fully exploited.
Indeed, besides the notable cases of the synchronisation of the cofluctuations across subjects watching the same movies and the peak-to-peak interval distribution (see \fig{inter_event_distribution_GSR}), most of the proposed edge-centric metrics are unaffected by time shuffling (\ie~they are static measures) and can thus be replicated by the i.i.d. null model.
\q{R3-slidingwindows}{Our mathematical derivations do not directly apply to sliding-window approaches since windowed correlations are changed by time shuffling, but it is important to remember that many common time-varying FC analysis pipelines have intermediate steps that alternately leverage and neglect temporal ordering~\citep{Lurie2020}.
For example, one might estimate sliding-window correlations (dynamic stage), apply k-means clustering to the resulting time-resolved FC matrices (static stage, since k-means ignores the temporal ordering of the windows), and then evaluate state properties such as dwell times and transition probabilities (dynamic stage).
While dynamic measures cannot be predicted from the static nFC, it is not unlikely that static null models could reproduce some of the static measures involved, \eg~the state patterns found via k-means, but not their transition probabilities.}

Let us note, however, that the role of null models in time-varying FC is a matter of current debate~\citep{Liegeois2021,Liegeois2017}, and not all features that can be explained by null models are clinically irrelevant or to be dismissed.
For example, using a small fraction of high-amplitude frames to approximate the nFC has been suggested as a way to compress the BOLD signal and alleviate the computational burden of analysing large fMRI datasets without compromising the prediction accuracy~\citep{Tagliazucchi2016}.
\q{R2-filtering}{Future studies may find other useful criteria for filtering the frames using the edge time series.}
As a final contribution, we have analytically shown that the predominance of a few frames in shaping the nFC is to be expected: the nFC captures the BOLD signal variance, which is a second order statistic and heavy-tailed (even if the BOLD signal were Gaussian).
Therefore, the nFC is necessarily shaped by a few tail events corresponding to large amplitude frames.
These frames determine the direction of maximum variance and thus the first principal components of the BOLD signal, forming the leading eigenvectors of the nFC.

In conclusion, we have laid out the mathematical foundations for the edge-centric FC analysis with the goal of informing future studies, in an interplay with empirical observations and simulations.
Future work could leverage this mathematical framework and focus on dynamic measures that cannot be easily explained by minimal null models like the one presented herein.

\ifarXiv
\else
\bibliography{bibliography}
\fi

\section{Methods}
\label{sec:methods}

\subsection{Definition of edge-centric FC}
Functional connectivity is defined as the magnitude of the statistical dependence between pairs of brain parcels~\citep{Friston1994functional}.
This dependence is typically estimated from their time series (here, the BOLD signal) using the Pearson correlation coefficient.
Let $N$ be the number of parcels, $T$ be the number of recorded frames, and $\mathbf{x}_i = [x_i(1),\ldots,x_i(T)]$ be the time series recorded from parcel $i$, with $1\leq i \leq N$.
The correlation between two parcels $i$ and $j$ can be computed as $r_{ij}=\frac{1}{T-1}\sum_{t}z_i(t) z_j(t)$, where $\mathbf{z}_i$ and $\mathbf{z}_j$ are their z-scored time series row vectors, \ie~$\mathbf{z}_i=\frac{\mathbf{x}_i-\mu_i}{\sigma_i}$ (with $\mu_i$ and $\sigma_i$ indicating the time-averaged mean and standard deviation).
Repeating this procedure for all pairs of parcels results in a node-by-node $(N\times N)$ correlation matrix $\mathbf{R}=[r_{ij}]$, which is an estimate of the (node-centric) functional connectivity.

The edge time series between two parcels $i$ and $j$ is the vector resulting from the element-wise product of $\mathbf{z}_i$ and $\mathbf{z}_j$, which encodes the magnitude of their cofluctuations over time:
\begin{equation} \label{eq:edge_ts}
    c_{ij}(t) \coloneqq z_i(t) z_j(t).
\end{equation}
We will denote the random variable (RV) associated with the edge time series $c_{ij}(t)$ with a capital letter, \ie~$C_{ij}(t)$.
The column vector of all the $N^2$ edge time series values at a given time $t$ can be reshaped into a $N\times N$ matrix that is an instantaneous estimate of the dynamic functional connectivity based on a single frame.
This matrix is symmetric since each edge time series is independent of the order of the node pair it involves, \ie~$c_{ij}(t)=c_{ji}(t),\forall i,j$.
For this reason, only the upper-triangular portion is typically computed, for a total of $N(N-1)/2$ edge time series instead of $N^2$.

It is also possible to go one step further and estimate the statistical dependence between each pair of edge time series, where each edge corresponds to a pair of parcels.
This fourth-order statistic results in a large $\frac{N(N-1)}{2}\times\frac{N(N-1)}{2}$ matrix named edge functional connectivity~(eFC)~\citep{Faskowitz2020}, with normalised entries defined as
\begin{equation}\label{eq:eFC_definition}
    \mathrm{eFC}_{jk,lm} \coloneqq \frac{\sum_{t}c_{jk}(t)\ c_{lm}(t)}{\sqrt{\sum_{t}c_{jk}(t)^2}\sqrt{\sum_{t}c_{lm}(t)^2}}.
\end{equation}

\subsection{Null hypothesis}
Let $\mathbf{z}$ be the $N\times T$ (parcels $\times$ frames) matrix of z-scored BOLD observations.
Since our goal is to derive the (dynamic) edge-centric properties from the (static) nFC matrix ($\mathbf{R}=[r_{ij}]$), we need to define a null hypothesis that discounts any temporal dependencies but retains the observed spatial correlations in $\mathbf{R}$.
A simple null hypothesis on the distribution of (the columns of) $\mathbf{z}$ that satisfies this criterion is $\mathbf{Z}(t)\sim \mathcal{N}(0,\mathbf{R})$, that is, i.i.d. multivariate Gaussian RV with a covariance matrix $\mathbf{R}$ matching the observed nFC.
If we denote the state of the system at time $t$ as the column vector $\mathbf{z}(t)=[z_1(t),\ldots,z_N(t)]^\intercal$, the null hypothesis simply states that $\mathbf{z}(t)$ is drawn from the same multivariate Gaussian distribution at each time $t$, independently of the other samples.

\subsection{Derivation of edge FC}

With capital letters denoting RVs, the expected product between two edge time series $C_{jk}(t)$ and $C_{lm}(t)$ at time $t$ is
\begin{align}\label{eq:ets_product}
    \E[C_{jk}(t)C_{lm}(t)] =& \E[Z_j(t) Z_k(t) Z_l(t) Z_m(t)] \nn\\
    =& \kappa(Z_j(t) Z_k(t) Z_l(t) Z_m(t)) \nn\\
    &+\E[Z_j(t) Z_k(t)]\E[Z_l(t) Z_m(t)] \nn\\
    &+\E[Z_j(t) Z_l(t)]\E[Z_k(t) Z_m(t)] \nn\\
    &+\E[Z_j(t) Z_m(t)]\E[Z_k(t) Z_l(t)],
\end{align}
which follows from the definition of the joint cumulant $\kappa(Z_j(t) Z_k(t) Z_l(t) Z_m(t))$, also noting that that the products involving the expectation of a single variable are equal to zero (\ie~$\E[Z_i(t)]=0$) since $\mathbf{z}_i$ are z-scored.
The expression of the expectation in terms of joint cumulants is sometimes referred to as moment-cumulants formula~\citep{Speed1983,Rosenblatt1985}. 
The joint cumulant is equal to zero for Gaussian RVs~\citep{Rosenblatt1985}, allowing a simplification of \eq{ets_product} known as Isserlis' theorem~\citep{Isserlis1918}:
\begin{align}\label{eq:isserlis}
    \E[C_{jk}(t)C_{lm}(t)] =& \E[Z_j(t) Z_k(t)]\E[Z_l(t) Z_m(t)] \nn\\
    +&\E[Z_j(t) Z_l(t)]\E[Z_k(t) Z_m(t)] \nn\\
    +&\E[Z_j(t) Z_m(t)]\E[Z_k(t) Z_l(t)].
\end{align}
If we additionally assume that $\mathbf{Z}(t)$ is ergodic (which does not preclude it from being a multivariate autoregressive process~\citep{Liegeois2017}), all the involved terms become independent of time and \eq{isserlis} further simplifies to
\begin{align}\label{eq:ets_product_stationary}
    \E[C_{jk}(t)C_{lm}(t)] =& r_{jk}r_{lm} + r_{jl}r_{km} + r_{jm}r_{kl}.
\end{align}
In a mathematical sense, the expectation is to be intended over the population (ensemble), whereas the eFC is computed from a single session (sample).
However, the ergodic assumption guarantees that sample estimates converge to the ensemble expectation as the number of time frames increases.
We can then obtain an estimate of the eFC by substituting \eq{ets_product_stationary} into \eq{eFC_definition}:
\begin{align}\label{eq:eFC_null}
    \mathrm{eFC}_{jk,lm} =& \frac{r_{jk}r_{lm} + r_{jl}r_{km} + r_{jm}r_{kl}}{\sqrt{1 + 2 {r_{jk}}^2}\sqrt{1 + 2 {r_{lm}}^2}}.
\end{align}
The i.i.d. multivariate Gaussian null model clearly satisfies the ergodic hypothesis since it has no memory.
However, note that neither the i.i.d. nor the Gaussian properties are necessary since the derivations in this section assume ergodicity with the only additional constraint that the joint cumulant is equal to zero.

\subsection{Derivation of edge communities}
\label{sec:methods_edge_communities}
In order to formalise the intuition that two edges $(jk)$ and $(j'k')$ with similar rows of the eFC are likely to be clustered together, let us define their distance $d_{jk,j'k'}$ as the $\ell^1$ norm of the difference between the corresponding
rows of the (unnormalised) eFC:
\begin{align}\label{eq:dist_edges}
    d_{jk,j'k'} \coloneqq \sum_{l,m=1}^N & \frac{1}{T}|\E\left[C_{jk}C_{lm}^\intercal\right] - \E\left[C_{j'k'}C_{lm}^\intercal\right]| \nn\\
    =\sum_{l,m=1}^N &| r_{jk}r_{lm} + r_{jl}r_{km} + r_{jm}r_{kl} \nn\\
        -& r_{j'k'}r_{lm} - r_{j'l}r_{k'm} - r_{j'm}r_{k'l}| \nn\\
    =\sum_{l,m=1}^N &| \mathbf{z}_j \mathbf{z}^\intercal_k \mathbf{z}_l \mathbf{z}^\intercal_m + \mathbf{z}_j \mathbf{z}^\intercal_l \mathbf{z}_k \mathbf{z}^\intercal_m + \mathbf{z}_j \mathbf{z}^\intercal_m \mathbf{z}_k \mathbf{z}^\intercal_l \nn\\
        -& \mathbf{z}_{j'} \mathbf{z}^\intercal_{k'} \mathbf{z}_l \mathbf{z}^\intercal_m - \mathbf{z}_{j'} \mathbf{z}^\intercal_l \mathbf{z}_{k'} \mathbf{z}^\intercal_m - \mathbf{z}_{j'} \mathbf{z}^\intercal_m \mathbf{z}_{k'} \mathbf{z}^\intercal_l| \nn\\
    =\sum_{l,m=1}^N &3|\mathbf{z}_j (\mathbf{z}^\intercal_l \mathbf{z}_m) \mathbf{z}^\intercal_k - \mathbf{z}_{j'} (\mathbf{z}^\intercal_l \mathbf{z}_m) \mathbf{z}^\intercal_{k'}|.
\end{align}
We now have the necessary ingredients to build a measure of similarity between the nodes, which can be used to predict the edge cluster similarity in~\citep{Faskowitz2020}.
There, the similarity between two nodes is measured as the frequency with which the corresponding edges are clustered together (having fixed the number of communities to $10$).
Instead of discrete assignments to $10$ communities, \eq{dist_edges} provides a continuous measure of the distance between two edges.
The distance between two nodes $i$ and $j$ can then be defined as the sum of the distances between the edges starting from $i$ and $j$ and reaching the same target: 
\begin{align}\label{eq:dist_nodes}
    d_{i,j} &= \sum_{k=1}^N d_{ik,jk} \nn\\
    &= \sum_{k,l,m} 3|(\mathbf{z}_i - \mathbf{z}_j)(\mathbf{z}^\intercal_l \mathbf{z}_m) \mathbf{z}^\intercal_k| \nn\\
    &\leq \lVert (\mathbf{z}_i - \mathbf{z}_j)\rVert \sum_{k,l,m} 3\lVert(\mathbf{z}^\intercal_l \mathbf{z}_m) \mathbf{z}^\intercal_k\rVert \nn\\
    &= \left(\sum_t (z_i(t)-z_j(t))^2\right)^\frac{1}{2} \sum_{k,l,m} 3\lVert(\mathbf{z}^\intercal_l \mathbf{z}_m) \mathbf{z}^\intercal_k\rVert \nn\\
    &=\! \left((T-1)(\mathrm{Var}[\mathbf{z}_i]\! +\! \mathrm{Var}[\mathbf{z}_j]\! -\! 2r_{ij})\right)^\frac{1}{2}\!\sum_{k,l,m} \! 3\lVert(\mathbf{z}^\intercal_l \mathbf{z}_m) \mathbf{z}^\intercal_k\rVert \nn\\
    &= \left(1-r_{ij}\right)^\frac{1}{2} \left[(2(T-1))^\frac{1}{2} \sum_{k,l,m} 3\lVert(\mathbf{z}^\intercal_l \mathbf{z}_m) \mathbf{z}^\intercal_k\rVert \right]\nn\\
    &\propto \left(1-r_{ij}\right)^\frac{1}{2},
\end{align}
where we used the Cauchy-Schwarz inequality and noted that the terms in the square brackets form a constant (independent of $i$ and $j$).
It is then apparent that the edge-cluster similarity~\citep{Faskowitz2020} between nodes $i$ and $j$ can be approximated by the nFC.
Once again, note that the i.i.d. Gaussian RV assumption can be relaxed since the derivations in this section are based on \eq{ets_product_stationary}, which requires ergodicity with the only additional constraint that the joint cumulant is equal to zero.

\subsection{Derivation of RSS from the BOLD signal}
Recalling the definition of the edge time series $c_{ij}(t)$ in \eq{edge_ts},
the RSS defined in~\citep{Esfahlani2020} can be approximated as the squared Euclidean norm of the z-scored BOLD signal, up to a constant factor:
\begin{align} \label{eq:RSS_as_Euclidean_approx}
    \mathrm{RSS}(t) &\coloneqq \sqrt{\sum_{i<j}{c_{ij}(t)}^2} \nn\\
    &=\sqrt{\frac{1}{2}\left(\sum_{i,j=1}^N {c_{ij}(t)}^2 - \sum_{i=1}^N{c_{ii}(t)}^2\right)} \nn\\
    &=\sqrt{\frac{1}{2}\left(\sum_{i,j=1}^N z_i(t)^2 z_j(t)^2 - \sum_{i=1}^N z_{i}(t)^4\right)} \nn\\
    &=\sqrt{\frac{1}{2}\left({\lVert \mathbf{z}(t)\rVert}^4-\sum_{i=1}^N z_{i}(t)^4\right)} \nn\\
    &\approx\frac{1}{\sqrt{2}}{\lVert \mathbf{z}(t)\rVert}^2.
\end{align}
The approximation does not rely on the i.i.d. assumption; it is valid under the Gaussian null hypothesis and, more generally, for distributions with finite kurtosis -- including fMRI data.
Under this assumption, ${\lVert \mathbf{z}(t)\rVert}^4$ dominates $\sum_i z_{i}(t)^4$ in \eq{RSS_as_Euclidean_approx}, as can be seen from the ratio of their (expected) values:
\begin{align}
    \frac{\E\left[\sum_i Z_{i}(t)^4\right]}{\E\left[{\lVert \mathbf{Z}(t)\rVert}^4\right]} \leq \frac{\sum_i\mathrm{Kurt}[Z_i(t)]}{N^2} \xrightarrow[N \to \infty]{} 0.
\end{align}

The approximation in \eq{RSS_as_Euclidean_approx} can be replaced by an exact equality if all the $N^2$ edge time series are included in the RSS definition (that is, all the $(i,j)$ tuples, rather than only the pairs with $i<j$):
\begin{align} \label{eq:RSS_as_Euclidean_exact}
    \mathrm{RSS}_\mathrm{all}(t) &\coloneqq \sqrt{\sum_{i,j}{c_{ij}(t)}^2} ={\lVert \mathbf{z}(t)\rVert}^2.
\end{align}

\subsection{Why are frames with the largest RSS most similar to the nFC?}
\label{sec:methods_eigenvectors}
Having rewritten the RSS as the squared Euclidean norm in \eq{RSS_as_Euclidean_approx}, we can more easily investigate the conditions underpinning the largest RSS fluctuations.
Let us withen the BOLD vector $\mathbf{Z}(t)$ to obtain the RV
\begin{align}
    \mathbf{W}(t) &\coloneqq \mathbf{R}^{-\frac{1}{2}} \mathbf{Z}(t)
\end{align}
and let
\begin{align}
    \mathbf{R} &= \mathbf{U}\mathbf{\Lambda} \mathbf{U}^\intercal
\end{align}
be the eigendecomposition of the correlation matrix $\mathbf{R}$, with $\mathbf{\Lambda}=\mathrm{diag}(\lambda_1,\ldots,\lambda_N)$ and $\mathbf{U}$ being the unitary matrix of eigenvectors of $\mathbf{R}$ (since $\mathbf{R}$ is symmetric).
Without loss of generality, assume that the eigenvalues are sorted in descending order, such that $\lambda_1$ is the largest eigenvalue and $\mathbf{u}_1$ is the corresponding leading eigenvector.
The RSS can be treated as a RV and rewritten in terms of $\mathbf{W}(t)$ and the eigenvector matrix $\mathbf{U}$:
\begin{align}
    \mathrm{RSS}(t) &\approx \frac{1}{\sqrt{2}}{\lVert \mathbf{Z}(t)\rVert}^2 = \frac{1}{\sqrt{2}} \mathbf{Z}(t)^\intercal \mathbf{Z}(t)  \nn\\
    &=\frac{1}{\sqrt{2}} \mathbf{W}(t)^\intercal(\mathbf{R}^\frac{1}{2})^\intercal (\mathbf{R}^\frac{1}{2}) \mathbf{W}(t) \nn\\
    &=\frac{1}{\sqrt{2}} (\mathbf{W}(t)^\intercal \mathbf{U})\mathbf{\Lambda}(\mathbf{U}^\intercal \mathbf{W}(t)) \nn\\
    &=\frac{1}{\sqrt{2}} \sum_i \lambda_i [(\mathbf{U}^\intercal \mathbf{W}(t))_i]^2 \nn\\
    &=\frac{1}{\sqrt{2}} \sum_i \lambda_i \langle \mathbf{u}_i,\mathbf{W}(t)\rangle^2 \label{eq:RSS_eigenvalues} \\
    &=\frac{1}{\sqrt{2}} \sum_i \lambda_i \lVert \mathbf{u}_i \rVert^2 \lVert \mathbf{W}(t)\rVert^2 \cos^2{\Theta_i(t)},
\end{align}
where $\mathbf{u}_i$ is the $i$-th eigenvector and $\Theta_i(t)$ is the RV representing the angle formed by the vectors $\mathbf{u}_i$ and $\mathbf{W}(t)$ at time $t$.
Also note that $\lVert \mathbf{u}_i \rVert^2=1$ because $\mathbf{U}$ is unitary.
For any realisations $\mathbf{w}(t)$ with squared norm $\lVert \mathbf{w}(t)\rVert^2$, an upper bound on the RSS is obtained as
\begin{align}\label{eq:RSS_upper_bound}
    \mathrm{RSS}(t) &\leq \frac{1}{\sqrt{2}}\lVert \mathbf{w}(t) \rVert^2 \max_{i}{\lambda_i} \sum_i\cos^2{\theta_i(t)} \nn\\
    &= \frac{1}{\sqrt{2}}{\lambda_\mathrm{max}} \lVert \mathbf{w}(t) \rVert^2,
\end{align}
noting that
\begin{align}
    \sum_i\cos^2{\theta_i(t)} &= \sum_i \frac{\langle \mathbf{u}_i,\mathbf{w}(t)\rangle^2}{\lVert \mathbf{u}_i \rVert^2 \lVert \mathbf{w}(t) \rVert^2} = \frac{\lVert \mathbf{U} \mathbf{w}(t) \rVert^2}{\lVert \mathbf{w}(t)   \rVert^2} = 1.
\end{align}
The upper bound is reached when $\theta_1(t')=0$ or $\theta_1(t')=\pi$, which implies that $\mathbf{w}(t')=c\,  \mathbf{u}_1$, where $c$ is a constant.
In other words, $\mathbf{w}(t')$ is aligned (parallel or antiparallel) with the leading eigenvector $\mathbf{u}_1$.
When this happens, the BOLD signal vector $\mathbf{z}(t')$ must also be aligned with $\mathbf{u}_1$:
\begin{align}\label{eq:BOLD_aligned_with_PC1}
    \mathbf{z}(t') &= \mathbf{R}^\frac{1}{2} \mathbf{w}(t') = \mathbf{R}^\frac{1}{2} c\,  \mathbf{u}_1 \nn\\
    &= \mathbf{U}\mathbf{\Lambda}^\frac{1}{2} \mathbf{U}^\intercal \mathbf{u}_1 = c\,  \lambda_1^\frac{1}{2} \mathbf{u}_1.
\end{align}
We can then refine our theoretical understanding of the RSS peaks: not only they occur when the Euclidean norm of the BOLD signal vector is large (as per \eq{RSS_as_Euclidean_approx}) but, most likely, when it is well aligned with the leading eigenvector of the static nFC (see \fig{PC_plots_GSR}b).
If the alignment were perfect at a frame $t'$, the instantaneous estimate of the nFC would be
\begin{align}\label{eq:ets_aligned_with_PC1}
    \mathbf{z}(t')\mathbf{z}(t')^\intercal &= c^2 \lambda_1 \mathbf{u}_1\mathbf{u}_1^\intercal,
\end{align}
\ie~an approximation of the nFC obtained from its leading eigenvector only (and independent of the sign of $c$).
This approximation would achieve an average similarity (Pearson correlation coefficient) of $r=0.69$ with the nFC over $100$ unrelated participants of the HCP dataset (while, in practice, the highest similarity achieved by the top frame was $r=0.53$).
However, the alignment with $\mathbf{u}_1$ need not be perfect: in general, large RSS values can be expected whenever the BOLD signal is a mixture of the top eigenvectors.
Additional principal components are expressed as more large-RSS frames are averaged, suggesting why the top $5\%$ frames alone are sufficient for an almost perfect reconstruction of the nFC~(\fig{nFC_similarity_GSR}).   
We have thus explained why cofluctuation patterns corresponding to frames with the largest RSS exhibit the highest similarity with the nFC.
These results are based on \eq{RSS_as_Euclidean_approx} and hold true under the assumption of finite kurtosis (which also applies in the specific case of the null hypothesis, \ie~for Gaussian variables).
The i.i.d. assumption is not required.

\subsection{Null distribution of the RSS}
The RSS can be written as a simple quadratic form $\mathrm{RSS}(t) = \frac{1}{\sqrt{2}}{\lVert \mathbf{Z}(t)\rVert}^2 = \frac{1}{\sqrt{2}} \mathbf{Z}(t)^\intercal \mathbf{Z}(t)$, which is known to follow a generalised $\chi^2$ distribution under the null hypothesis of Gaussian variables~\citep{Mathai1992}.
The weights of the non-central chi-square components are proportional to the eigenvalues of the nFC matrix, \ie~$\frac{\lambda_1}{\sqrt{2}},\ldots,\frac{\lambda_N}{\sqrt{2}}$.
Another characterisation of this distribution is provided by \eq{RSS_eigenvalues}: under the null hypothesis, the inner product $\langle \mathbf{u}_i,\mathbf{W}(t)\rangle$ follows a normal Gaussian distribution since $\mathbf{W}(t)\sim\mathcal{N}(0,1)$ and $\mathbf{U}$ is unitary.
Therefore, $\langle \mathbf{u}_i,\mathbf{W}(t)\rangle^2$ follows a $\chi^2$ distribution and each term $\frac{\lambda_i}{\sqrt{2}} \langle \mathbf{u}_i,\mathbf{W}(t)\rangle^2$ in \eq{RSS_eigenvalues} follows a Gamma($k=\frac{1}{2},\theta=\sqrt{2}\lambda_i$) distribution. 
The RSS is thus obtained as a sum of $N$ independent Gamma-distributed RVs, each associated with one eigenvalue of the nFC.
The tail of the RSS is best approximated by the RVs associated with the largest eigenvalues (which have the largest mean and variance), while including smaller eigenvalues provides an increasingly fuller characterisation of the whole distribution (\fig{CDFCDF_plots}b).
The mean and variance of the RSS can be readily obtained from the properties of the Gamma distribution:
\begin{align}\label{eq:RSS_expectation}
    \E[\mathrm{RSS}] &= \frac{1}{\sqrt{2}} \sum_i \lambda_i=\frac{N}{\sqrt{2}} \\
    \mathrm{Var}[\mathrm{RSS}] &= \sum_i \lambda_i^2.
\end{align}
Higher moments of the RSS null distribution can be derived from its moment-generating function:
\begin{align} \label{eq:RSS_mgf}
    M_\mathrm{RSS}(s)=\prod_i (1-\sqrt{2}\lambda_i s)^{-\frac{1}{2}}.
\end{align}

\subsection{Why are large RSS fluctuations present in many datasets?}
The moment-generating function in \eq{RSS_mgf} can be employed to show that the RSS is subexponential under the null hypothesis, which explains its heavy tail and the consequent large events~\citep{Foss2013,Filiasi2014}.
Specifically, the subexponential feature of the null RSS follows from the sufficient condition
\begin{align} \label{eq:sub_gaussian}
    M_{\mathrm{RSS}-\E[\mathrm{RSS}]}(s) =& \prod_i (1-\sqrt{2}\lambda_i s)^{-\frac{1}{2}} \exp{-\frac{\lambda_i s}{\sqrt{2}}} \nn\\
    \leq& \prod_i \exp{\lambda_i^2 s^2} = \exp{s^2 \sum_i \lambda_i^2}, \\
    &\forall |s|\leq (4\lambda_\mathrm{max})^{-1} \nn.
\end{align}
However, we can expect this behaviour under the more general hypothesis that the z-scored BOLD signal is sub-Gaussian, \ie~its tail decays at least as fast as that of a Gaussian RV (including, for example, any uniformly-bounded RVs).
The reason is that the square of a sub-Gaussian RV is sub-exponential, and the sum of independent subexponential RVs is also subexponential.
Therefore, being the RSS closely approximated by a sum of squared RVs (as per \eq{RSS_as_Euclidean_approx}), extreme events are to be expected under the general sub-Gaussian assumption for the BOLD signal, which offers an explanation for the large RSS fluctuations observed in most fMRI datasets.

\subsection{How do nFC modules influence the edge cofluctuations?}
Interestingly, \citet{Pope2021} have recently reported a connection between the presence of structural modules and the occurrence of large events in the edge cofluctuations (RSS).
Insofar as structural and functional modules are in agreement~\citep{Honey2007,Cabral2011}, we can explain these findings based on the nFC spectrum.
How do functional modules shape the eigenspectrum of the nFC?
In the ideal case of a block-diagonal matrix (with zeroes outside the blocks), the sum of the eigenvalues corresponding to each block coincides with the block size (since the diagonal elements are all ones and the trace is preserved under diagonalisation).
As such, the largest eigenvalue is bounded by the size of the largest block, \ie~larger functional modules allow for larger eigenvalues.
In turn, large eigenvalues underpin the high-amplitude cofluctuations, as shown in \secRef{methods_eigenvectors}.
Therefore, if the size of the modules is reduced via randomisation of the structural connectivity as in~\citep[SI Fig. 3]{Pope2021}, the expected magnitude of the RSS cofluctuations will drop according to \eq{RSS_upper_bound}.
This offers a mathematical explanation for the lower RSS event count when the modular structure is disrupted.


\subsection{A null model for binary edge time series}
Consider two parcels $j$ and $k$ with z-scored BOLD activity $Z_j(t)$ and $Z_k(t)$ and correlation coefficient $r_{jk}$ as defined by the nFC matrix.
Under the static Gaussian null hypothesis, $Z_k(t)\sim \mathcal{N}(0,1)$ and
\begin{align}\label{eq:correlated_gaussian_rvs_with_unit_variance}
    Z_j(t) = r_{kj} Z_k(t) + \sqrt{1-r_{kj}^2} V_j\ ,
\end{align}
where $V_j\sim \mathcal{N}(0,1)$, and the square root coefficient ensures unit variance for $Z_j(t)$ since both parcels are z-scored.
The edge time series corresponding to the two parcels is $C_{jk}(t)=Z_j(t)Z_k(t)$, as per \eq{edge_ts}.
By definition, the binary edge time series $\overline{C_{jk}(t)}$ is equal to $1$ when the cofluctuations between $j$ and $k$ are positive, \ie~when $Z_j(t)$ and $Z_k(t)$ have the same sign.
In other words, $\overline{C_{jk}(t)}$ is a Bernoulli($p_{jk}$) RV with probability
\begin{align}\label{eq:bernoulli_p_methods}
    p_{jk} &= P[\overline{C_{jk}(t)}>0] \nn\\
    &= P[Z_j(t)Z_k(t)>0] \nn\\
    &= 2 P[(Z_k(t)>0)\cap (Z_j(t)>0)] \nn\\
    &= 2 P\left[(Z_k(t)>0)\cap (r_{kj} Z_k(t) + \sqrt{1-r_{kj}^2} V_j>0)\right] \nn\\
    &= 2 P\left[(Z_k(t)>0)\cap \left(\frac{V_j}{Z_k(t)}>-\frac{r_{kj}}{\sqrt{1-r_{kj}^2}}\right)\right] \nn\\
    &= 2 P[A],
\end{align}
where $A$ is the event $(Z_k(t)>0)\cap \left(\frac{V_j}{Z_k(t)}>-\frac{r_{kj}}{\sqrt{1-r_{kj}^2}}\right)$.
From a geometric perspective, $A$ is satisfied by any vector $(Z_k,V_j)$ whose polar angle is between $\pi /2$ and $\arctan(-\frac{r_{jk}}{\sqrt{1-r_{kj}^2}})$.
Therefore,
\begin{align}
    p_{jk} &= 2 P[A] \nn\\
    &= 2 \left(\frac{\frac{\pi}{2}-\arctan(-\frac{r_{jk}}{\sqrt{1-r_{kj}^2}})}{2\pi}\right) \nn\\
    &=\frac{1}{2}+\frac{\arcsin(r_{jk})}{\pi}.
\end{align}
In conclusion, under the null hypothesis, the binary edge time series $\overline{{C}_{jk}(t)}$ can be modelled as a Bernoulli($p_{jk}$) RV with $p_{jk}=\frac{1}{2}+\frac{\arcsin(r_{jk})}{\pi}$.

\subsection{Relationship with coactivation patterns (CAPs)}
Here, we will focus on explaining a well-known finding in the CAPs literature: choosing a seed node, \citet{Liu2013} observed a strong correlation between the BOLD activity at frames where the seed is highly active and the corresponding seed-based FC.
Note that the correlation is measured between two $N$-dimensional vectors (where $N$ is the number of nodes): the first vector is the BOLD signal and the second vector is the column of the nFC matrix that corresponds to the chosen seed node.
In the following, we will show that the i.i.d. Gaussian null model can predict the observed correlation.
Let $k$ be the seed node and $\mathbf{z}_k$ its z-scored BOLD time series.
As usual, the associated random process is denoted with the capital letter $Z_k(t)$. 
If we condition on a specific value of the seed node, say $z^*_k\coloneqq Z_k(t^*)$, the BOLD activity vector at the corresponding time $t^*$ is expected to align with the k-th column of the nFC (denoted as $\hat{\mathbf{R}}_{k}$):
\begin{align}\label{eq:CAPS_expected_Z}
    \E[\mathbf{Z}(t^*)] = \E[\mathbf{Z}(t)|Z_k(t)=z^*_k] = z^*_k\:\hat{\mathbf{R}}_{k}.
\end{align}
This is an elementary property of multivariate Gaussian RVs that follows directly from \eq{correlated_gaussian_rvs_with_unit_variance}:
\begin{align}
    \E[Z_j(t^*)] &= \E[Z_j(t)|Z_k(t)=z^*_k] \nn\\
    &= \E\left[r_{kj} Z_k(t) + \sqrt{1-r_{kj}^2} V_j|Z_k(t)=z^*_k\right] \nn\\
    &= z^*_k\:r_{kj},\;\forall j=1,\ldots,N.
\end{align}
Perhaps less intuitively, the conditional correlation between the BOLD vector $\mathbf{Z}(t^*)$ and the k-th column of the sample nFC ($\hat{\mathbf{R}}_{k}$) is also proportional to the seed value $z^*_k$.
In order to prove it, let us first compute their expected inner product:
\begin{align}
    \E[\langle \mathbf{Z}(t^*),\hat{\mathbf{R}}_{k}\rangle]
    &=\E\left[\sum_j^N Z_j(t^*)\frac{1}{T-1}\sum_t^T Z_j(t)Z_k(t)\right] \nn\\
    &\approx \sum_{j} \E\left[Z_j(t^*)\right]\frac{1}{T-1}\sum_{t\neq t^*}\E\left[Z_j(t)Z_k(t)\right] \nn\\
    &= \sum_{j} z^*_k r_{kj}\;\frac{1}{T-1}\sum_{t\neq t^*}r_{kj} \nn\\
    &= z^*_k\:{\lVert \mathbf{R}_k\rVert}^2.
\end{align}
Similarly, the covariance can be approximated as
\begin{align}
    \mathrm{Cov}[\mathbf{Z}(t^*),\hat{\mathbf{R}}_{k}] &= \E\left[\frac{\langle \mathbf{Z}(t^*),\hat{\mathbf{R}}_{k}\rangle}{N-1}-\frac{\sum_{j} Z_j(t^*) \sum_{l} \hat{r}_{kl}}{N(N-1)}\right] \nn\\
    &\approx z^*_k\left(\frac{{\lVert \mathbf{R}_k\rVert}^2}{N}-\frac{\sum_{j\neq l} r_{kj}r_{kl}}{N(N-1)}\right)
\end{align}
The expected sample covariance is directly proportional to $z^*_k$ and thus peaks at frames with the highest activity of the seed node $k$.
This remains true after normalising the covariance to obtain the correlation coefficient, as shown in \fig{CAPs_similarity_GSR}a.

\subsection{Human Connectome Project fMRI Dataset}
This study used openly-available and independently-acquired resting-state fMRI (rsfMRI) data from the Human Connectome Project (HCP) S1200 release~\citep{VanEssen2013}.
In particular, we used the ``$100$ unrelated subjects'' dataset: a subset of $100$ non-twins adult participants which were pre-selected by the HCP coordinators ($54\%$ female; mean age $=29.11 \pm 3.67$ years; age range, $22$--$36$ years).
The HCP study was approved by the Washington University Institutional Review Board, and informed consent was obtained from all participants.
All subjects were scanned on a customized Siemens 3T ``Connectome Skyra'' with a $32$-channel head coil, housed at Washington University in St. Louis.
rsfMRI data was acquired in four runs of $15$ minutes over a $2$-day period, with eyes open and relaxed fixation on a projected bright cross-hair on a dark background (presented in a darkened room).
Resting state images were collected with the following parameters: gradient-echo EPI sequence, run duration = $14$:$33$ min, TR = $720$ ms,  TE = $33.1$ ms,  flip  angle = $52$°, FOV = $208$x$180$ mm (RO x PE), matrix = $104$x$90$ (RO x PE), slice thickness = $2$ mm, $2$-mm isotropic voxel resolution, multi-band factor = $8$, echo spacing = $0.58$ ms, BW = $2290$ Hz/Px).

\subsection{Pre-processing and ICA-FIX denoising}
Functional images in the HCP dataset were minimally pre-processed according to the pipeline described in~\citep{Glasser2013}.
In short, the data was corrected for gradient distortion, susceptibility distortion and motion and then aligned to a corresponding T1-weighted image with one spline interpolation step.
This volume was further corrected for intensity bias, normalised to a mean of $10000$, projected to the 32k\_fs\_LR mesh (excluding outliers), and aligned to a common space using a multi-modal surface registration.

In addition, the preprocessed rsfMRI data was cleaned of structured noise through a process that pairs independent component analysis (MELODIC) with FIX to automatically remove non-neural spatiotemporal components (trained on $25$ hand-labeled HCP subjects).
The FIX approach and initial results of classification accuracy are detailed in~\citep{Salimi-Khorshidi2014}, and the effects of the ICA + FIX  cleanup (and optimal methods to remove the artefactual components from the data) are evaluated in detail in~\citep{Griffanti2014}.
The cleaning pipeline is described more comprehensively in the HCP S1200 release reference manual (https://humanconnectome.org/study/hcp-young-adult/document/1200-subjects-data-release/) and the preprocessing and the cleaning scripts are openly available on Github (https://github.com/Washington-University/HCPpipelines).
The resulting ICA-FIX denoised rfMRI grayordinate surface timeseries are available as CIFTI files following the naming pattern: *REST{1,2}\_{LR,RL}\_Atlas\_MSMAll\_hp2000\_clean.dtseries.nii.

The Schaefer200 parcellation was used to define $200$ areas on the cerebral cortex~\citep{Schaefer2018}.
This functional parcellation was designed to optimise both local gradient and global similarity measures of the fMRI signal and is openly available in `32k fs LR' space for the HCP dataset.
The nodes are mapped to the Yeo canonical functional networks~\citep{Yeo2011}.
The parcellated data was analysed both before and after regressing the global signal.
The theoretical derivations and predictions hold and perform equally well in both cases, and we report any significant differences when they occur.
Unless otherwise stated, the GSR results are shown in the figures since they are more directly comparable to those published in~\citep{Esfahlani2020} and~\citep{Faskowitz2020}, noting in particular that GSR was performed in~\citep{Esfahlani2020}.
Despite the ICA-FIX preprocessing pipeline used here is entirely different from those employed in~\citep{Esfahlani2020} and~\citep{Faskowitz2020}, our results are in excellent agreement with the previously published ones.

\ifarXiv
\else
\fi

\section{Data and code availability}
\label{sec:data_availability}
The imaging data from the Human Connectome Project are publicly available and can be accessed after signing a data use agreement at https://db.humanconnectome.org.
Source data are provided with this paper.

\section{Code availability}
The analysis was performed with MATLAB (MathWorks, Inc., version 2020b) and the code is made freely available on Github for reproducibility (\url{https://github.com/LNov/eFC}).
A permanent record is also made available on Zenodo (\url{https://zenodo.org/record/6238564}).

\ifarXiv
\begin{acknowledgments}
\else
\acknowledgments
\fi
Data used in the preparation of this work was obtained from the MGH-USC Human Connectome Project (HCP) database (https://ida.loni.usc.edu/login.jsp). The HCP project is supported by the National Institute of Dental and Craniofacial Research (NIDCR), the National Institute of Mental Health (NIMH) and the National Institute of Neurological Disorders and Stroke (NINDS).
L.N. is funded by the Australian Research Council (Ref: DP200100757).
A.R. is funded by the Australian Research Council (Refs:  DE170100128  and  DP200100757) and Australian National Health and Medical Research Council Investigator Grant (Ref: 1194910).
A.R. is affiliated with The Wellcome Centre for Human Neuroimaging supported by core funding from Wellcome [203147/Z/16/Z].
A.R. is a CIFAR Azrieli Global Scholar in the Brain, Mind \& Consciousness Program.
We thank Joseph Lizier, Ben Fulcher, James Mac Shine, and Andrew Zalesky for their feedback on an earlier draft of this manuscript.
\ifarXiv
\end{acknowledgments}
\else
\fi

\ifarXiv
\section*{Author Contributions}
\else
\authorcontributions
\fi
Leonardo Novelli: Conceptualization; Data curation; Formal analysis; Investigation; Software; Visualization; Writing --- original draft.
Adeel Razi: Conceptualization; Funding acquisition; Supervision; Writing --- review \& editing.


\nolinenumbers

\ifarXiv

%

\else
\fi

\newcommand{\beginsupplement}{
    \renewcommand{\thepage}{S\arabic{page}} 
    \renewcommand{\thesection}{S\arabic{section}}
    \renewcommand{\theequation}{S\arabic{equation}}
    \renewcommand{\thetable}{S\arabic{table}}  
    \renewcommand{\thefigure}{S\arabic{figure}}
    \setcounter{page}{1}
    \setcounter{section}{1}
    \setcounter{table}{0}
    \setcounter{figure}{0}
    \setcounter{equation}{0}
    \newcounter{SIfig}
    \renewcommand{\theSIfig}{S\arabic{SIfig}}
    }

~
\onecolumngrid
\section*{Supplementary Material}
\FloatBarrier 
~

\beginsupplement

\begin{figure*}
    \ifarXiv\includegraphics[width=0.9\textwidth]{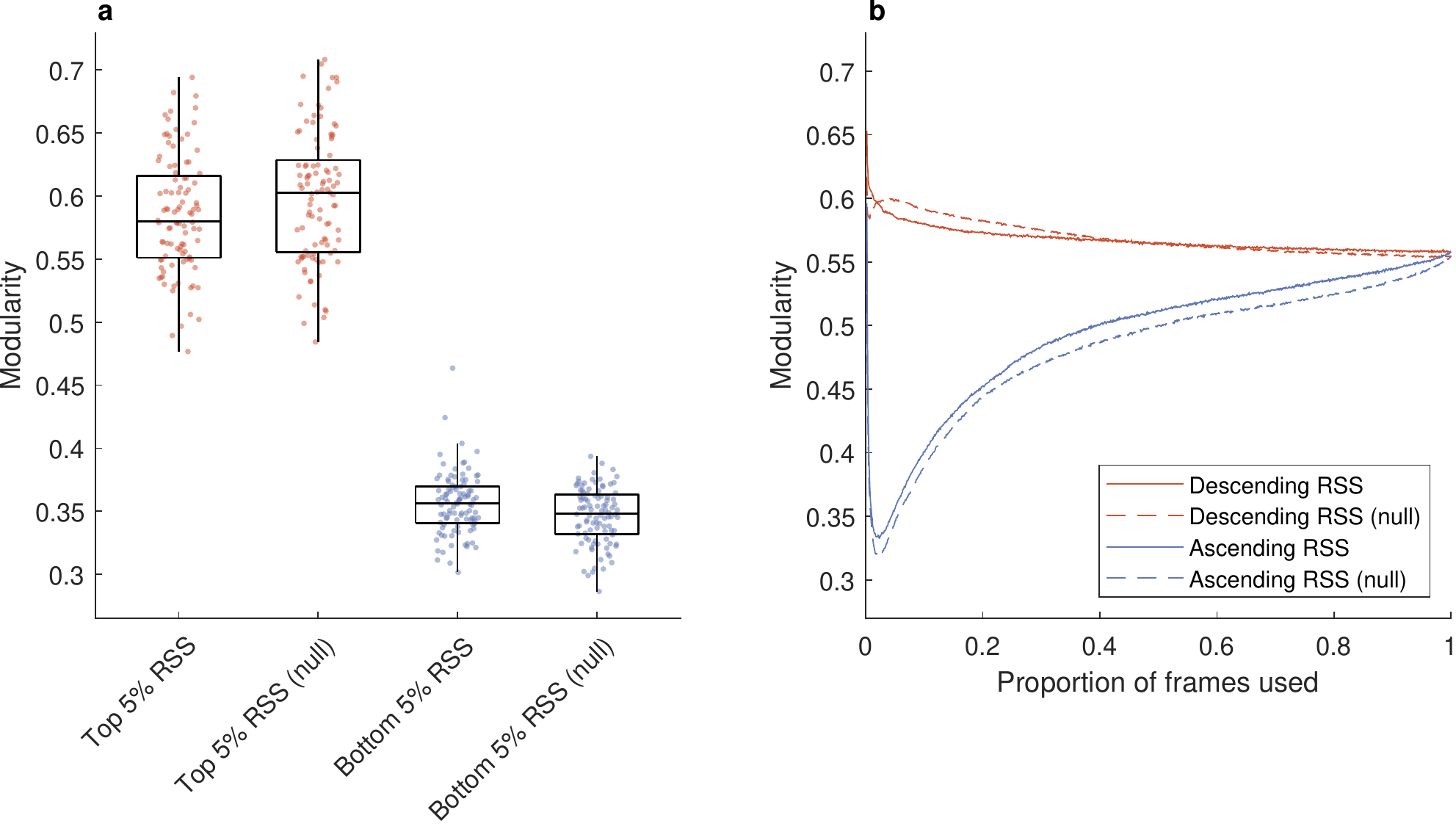}
    \else\centering\includegraphics[width=0.9\textwidth]{modularity_GSR}
    \fi
    \refstepcounter{SIfig}\label{fig:modularity_GSR}
    \caption{
        The static Gaussian null model can reproduce the empirical network modularity, which can be interpreted as a measure of the segregation between the network systems.
        The employed $q^*$ variant of modularity is well suited for use with correlation matrices~\cite{Rubinov2011}.
        \textbf{a)} The empirical results match the finding published in \cite[Fig. 1E]{Esfahlani2020}: the networks estimated using the top $5\%$ of frames exhibit much higher modularity than those estimated using the bottom $5\%$ of frames.
        The gap is accurately replicated by the null model.
        Each point corresponds to one of $100$ unrelated subjects from the Human Connectome Project (HCP) dataset, with boxes indicating the quartiles and whiskers length specified as $1.5$ times the interquartile range.
        \textbf{b)} The same results hold more generally when the frames are ordered according to the corresponding RSS amplitude, either in descending or ascending order.
        Here, the curves represent the average over $100$ subjects.
        }
\end{figure*}

\begin{figure*}
    \ifarXiv\includegraphics[width=\textwidth]{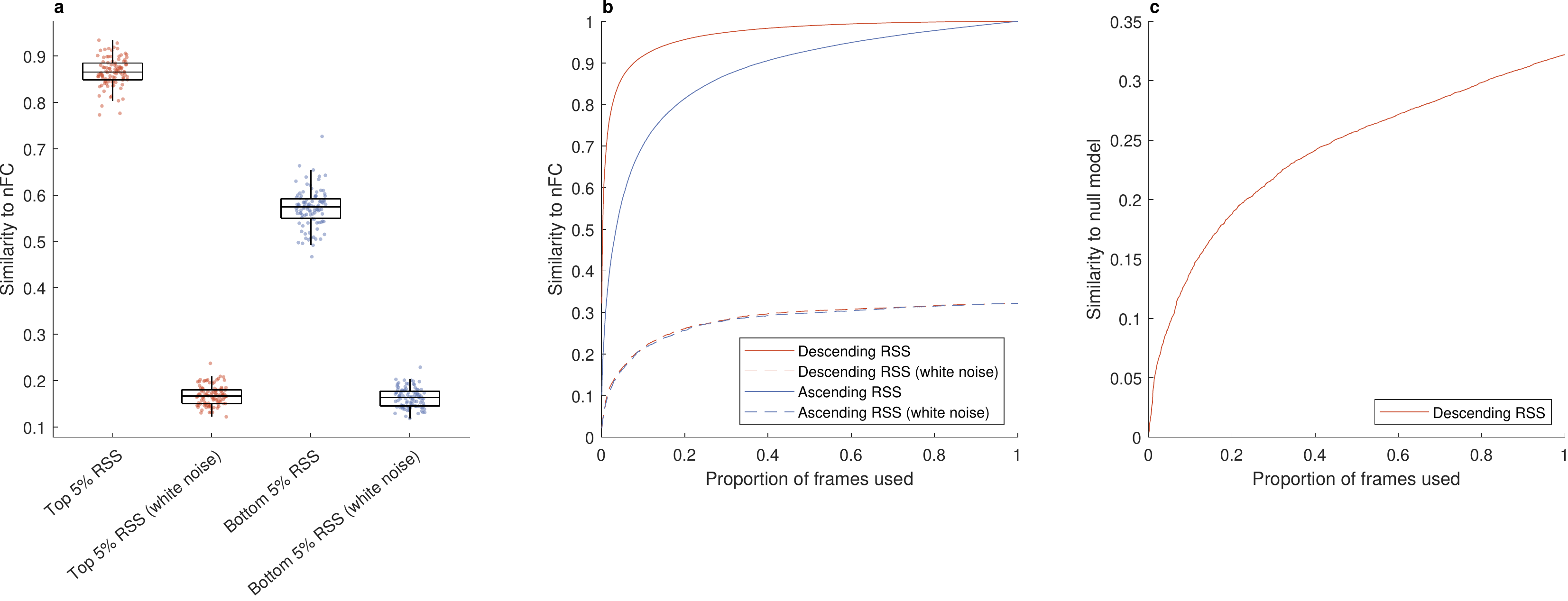}
    \else\centering\includegraphics[width=\textwidth]{nFC_similarity_white_noise_GSR}
    \fi
    \refstepcounter{SIfig}\label{fig:nFC_similarity_white_noise_GSR}
    \caption{
        Accounting for spatial correlations is essential to reproduce the empirical edge-centric findings.
        To prove that, we repeated the analysis reported in \fig{nFC_similarity_GSR} but ignored the spatial correlations in addition to the temporal ones.
        As a result, the differences between frames exhibiting high and low cofluctuation magnitudes (RSS) vanished.
        \textbf{a)} When using spatially uncorrelated white noise, the small fraction of frames exhibiting the largest RSS is no longer able explain most of the nFC variance (unlike the null model presented in the main text, which accounts for the observed spatial correlations encoded in the nFC matrix).
        The similarity is computed as the Pearson correlation coefficient between the nFC and the average FC estimated from the top and bottom $5$\% of the total frames.
        Each point corresponds to one of $100$ unrelated subjects from the Human Connectome Project (HCP) dataset, with boxes indicating the quartiles and whiskers length specified as $1.5$ times the interquartile range.
        \textbf{b)} The same inability is shown more generally when the frames are ordered according to the corresponding RSS amplitude, either in descending or ascending order.
        Here, the curves represent the average similarity over $100$ subjects.
        \textbf{c)} Unlike the spatially-correlated null model, the top frames generated by Gaussian white noise do not exhibit high similarity to the top frames of the real HCP data.
    }
\end{figure*}

\begin{figure*}
    \ifarXiv\includegraphics[width=\textwidth]{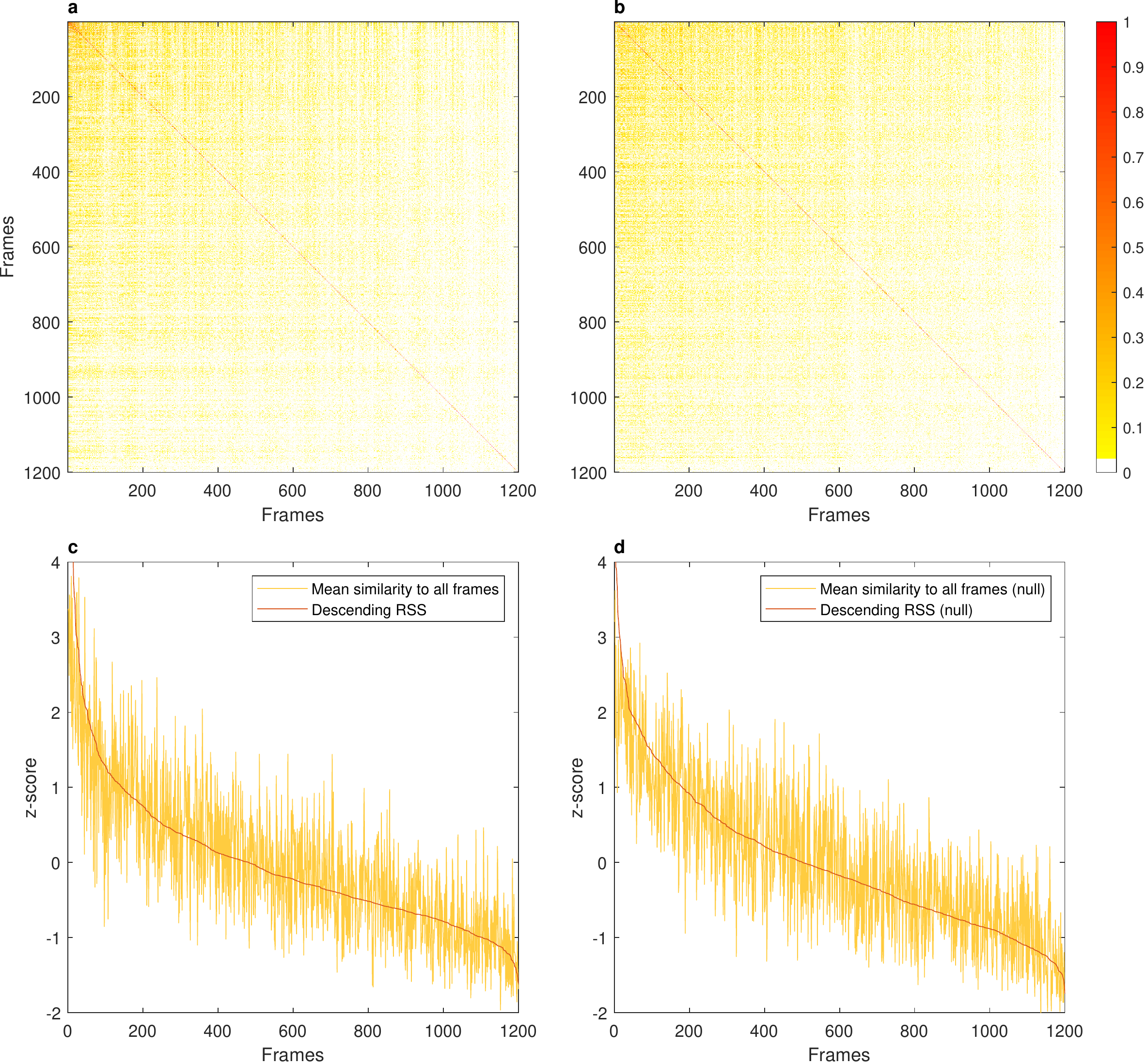}
    \else\centering\includegraphics[width=\textwidth]{similarity_to_all_frames_vs_descending_RSS_subj5}
    \fi
    \refstepcounter{SIfig}\label{fig:similarity_to_all_frames_vs_descending_RSS_subj5}
    \caption{
        Relationship between RSS magnitude and average similarity to all frames.
        \textbf{a)} Correlation between all pairs of cofluctuation patterns for a single HCP subject (ID 101915).
        Frames are arranged in descending order of RSS magnitude so that high-RSS frames are in the top-left corner.
        This is akin to rearranging the rows and columns of the FC dynamics matrix used in sliding-window approaches to time-varying FC~\cite{Hansen2015}.
        \textbf{b)} Corresponding null model results.
        \textbf{c)} Averaging the columns of the similarity matrix in panel \textbf{a} yields the mean similarity to all frames.
        The similarity decreases with the RSS, which is plotted in red.
        \textbf{d)} Corresponding null model results.
    }
\end{figure*}

\begin{figure*}
    \ifarXiv\includegraphics[width=\textwidth]{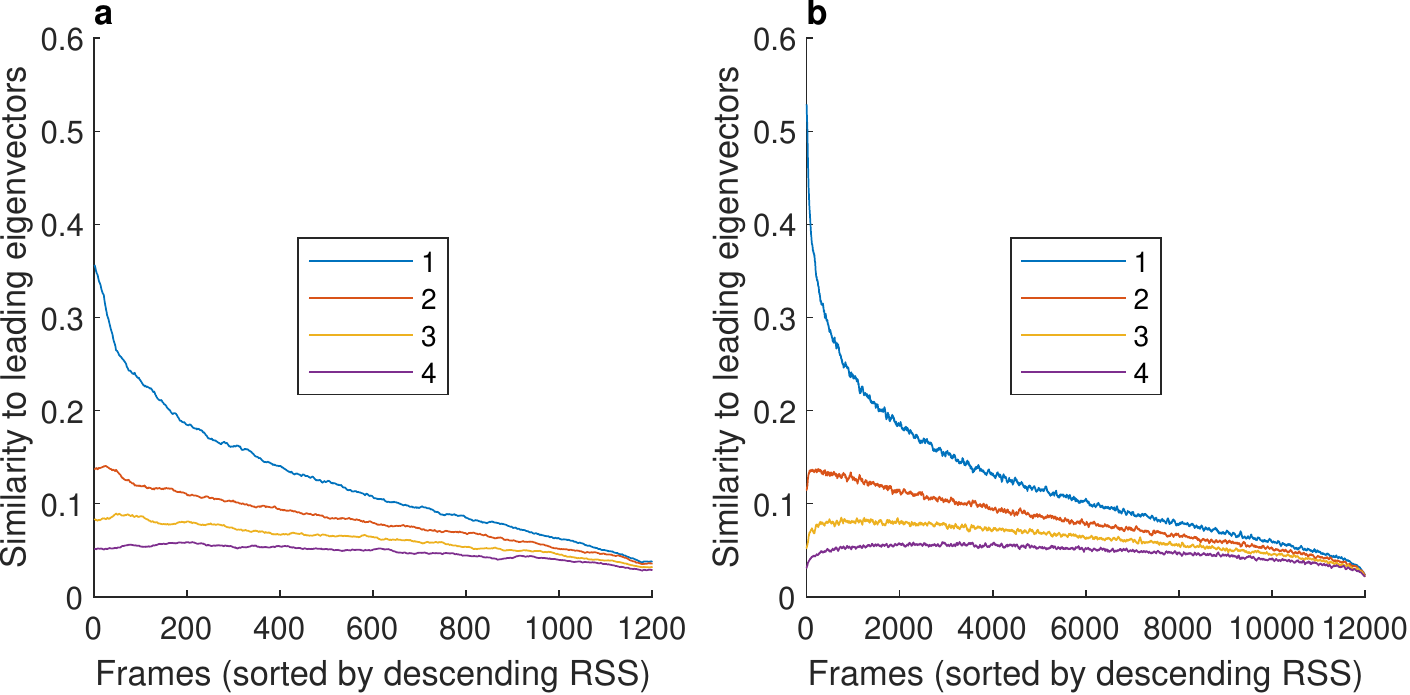}
    \else\centering
    \includegraphics[width=\textwidth]{eigenvectors_vs_RSS_GSR_1200vs12000}
    \fi
    \refstepcounter{SIfig}\label{fig:eigenvectors_vs_RSS_GSR_moreframes}
    \caption{
        Frames corresponding to large RSS values exhibit high similarity to the leading nFC eigenvectors.
        This similarity increases with the number of frames, as shown by comparing the empirical and synthetic results.
        \textbf{a}) Empirical results using \num{1200} frames available in the HCP dataset, as in Fig. 5b in the main text.
        The similarity is measured as the Pearson correlation between the FC estimate from a single frame and each of the estimates from the the four leading nFC eigenvectors.
        \textbf{b}) The same analysis was repeated on \num{10} times longer synthetic data (i.e. \num{12000} frames compared to previously analysed \num{1200} frames) generated for each HCP subject using their empirical nFC as input to the null model.
        The empirical nFC of a subject was set as the covariance matrix of a multivariate Gaussian process, which was sampled to obtain a longer synthetic scan for that subject.
    }
\end{figure*}

\begin{figure*}
    \ifarXiv\includegraphics[width=\textwidth]{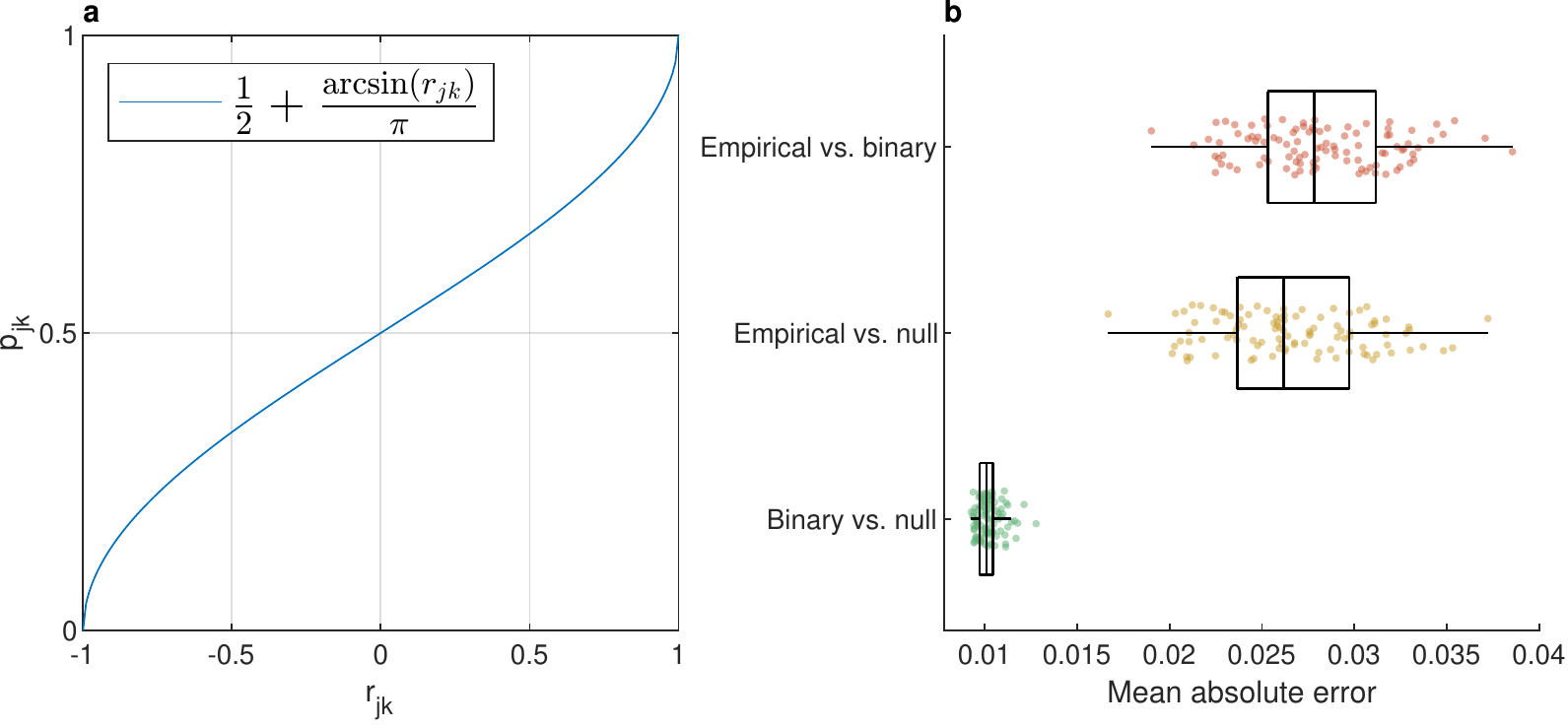}
    \else\centering\includegraphics[width=\textwidth]{bernoulli_prediction_GSR}
    \fi
    \refstepcounter{SIfig}\label{fig:bernoulli_prediction_GSR}
    \caption{
        Binary edge time series properties.
        \textbf{a)} Under the static Gaussian null model, the binary edge time series can be modelled as Bernoulli($p_{jk}$) random variables, where the success probability $p_{jk}$ is a function of the Pearson correlation coefficient $r_{jk}$.
        The plot confirms and extends the intuition that $p_{jk}=0$ for anticorrelated variables (parcels) and $p_{jk}=1$ for perfectly correlated ones, crossing $p_{jk}=0.5$ for uncorrelated variables.
        \textbf{b)} Averaging the binary edge time series over time yields a good approximation of the empirical nFC matrix computed on the HCP dataset, as measured by the mean absolute error (top box-whiskers plot, red dots).
        A similar error is obtained when comparing the nFC with the null model prediction produced using the function plotted in panel \textbf{a} (middle box-whiskers plot, yellow dots).
        Comparing the null prediction to the average binary time series results in an even lower mean error, validating the analytic derivation (bottom box-whiskers plot, green dots).
        Note that the empirical nFC must be rescaled to the $[0,1]$ interval to allow for a direct comparison with the average binary edge time series and the analytic null probability.
        Each point corresponds to one of $100$ unrelated subjects from the HCP, with boxes indicating the quartiles and whiskers length specified as $1.5$ times the interquartile range.
        }
\end{figure*}

\begin{figure*}
    \ifarXiv\includegraphics[width=\textwidth]{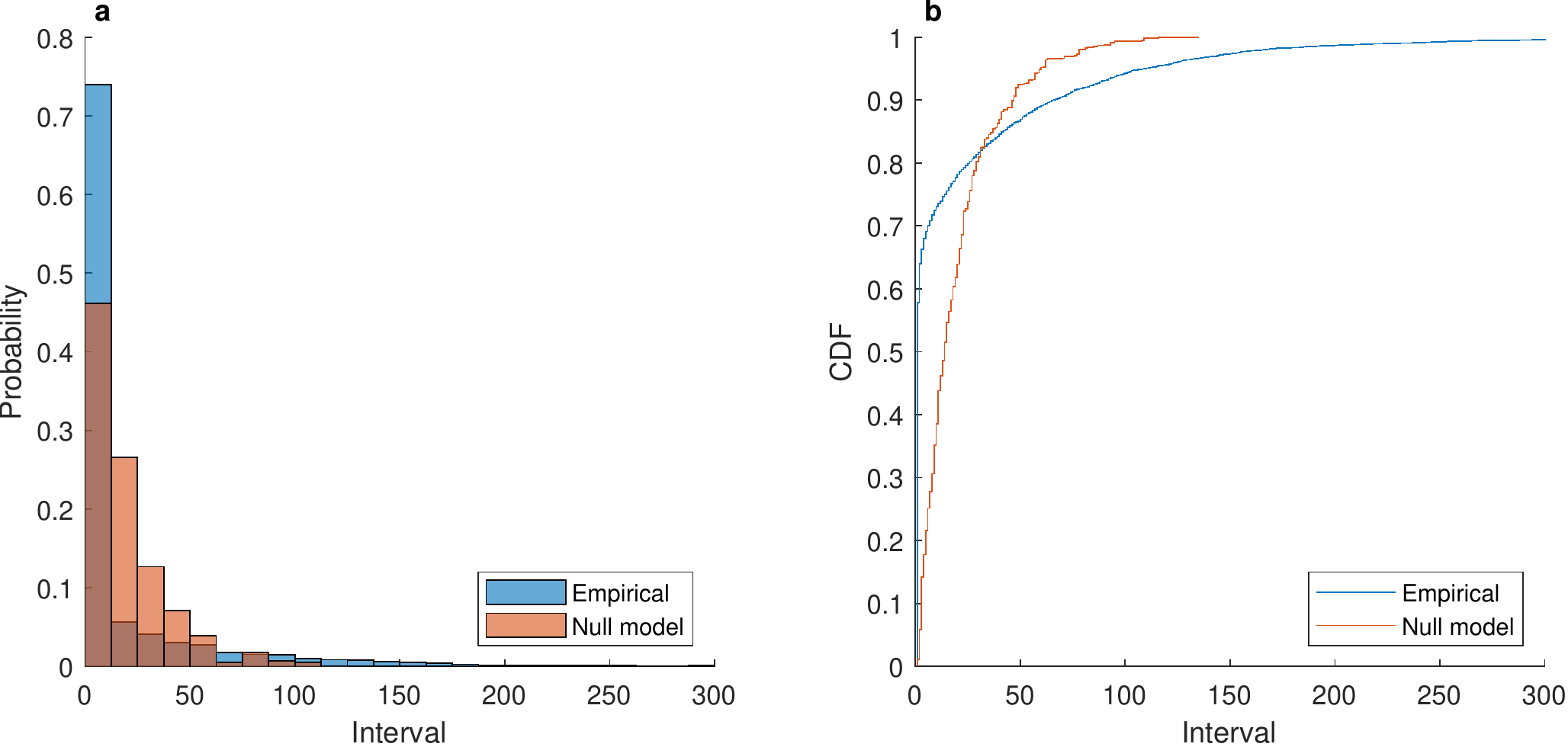}
    \else\centering\includegraphics[width=\textwidth]{inter_event_distribution_GSR}
    \fi
    \refstepcounter{SIfig}\label{fig:inter_event_distribution_GSR}
    \caption{
        Distribution of time intervals between high-amplitude cofluctuations, computed across $100$ unrelated HCP participants.
        The relative probability and cumulative distribution function (CDF) are reported in panels \textbf{a} and \textbf{b}, respectively.
        The root-sum-of-squares (RSS) of the edge time series was computed at each time step and the intervals between large realisations (those above the $95$th percentile) were computed.
        The null model only captures the observed spatial correlations and not the temporal ones; therefore, it cannot reproduce the empirical distribution of the intervals.
        }
\end{figure*}

\end{document}